\documentclass[fleqn,usenatbib]{mnras}

\usepackage{graphicx}	
\usepackage{amsmath}	
\usepackage{amssymb}	
\usepackage{mathtools}
\usepackage{adjustbox}
\usepackage{threeparttable}
\usepackage{tabto}
\usepackage{float}
\usepackage{color}
\usepackage{graphics}
\usepackage{adjustbox}
\usepackage[usenames,dvipsnames]{pstricks}
\usepackage{epsfig}
\usepackage{pst-grad} 
\usepackage{pst-plot} 
\usepackage{times,pstricks,pst-node,pst-coil}
\usepackage{multicol} 
\usepackage{multirow} 
\usepackage{indentfirst} 
\usepackage{natbib} 
\usepackage{threeparttable}
\usepackage{longtable}
\usepackage{pdflscape}

\newcommand{\Lsun}{L$_{\odot}$}
\newcommand{\Msun}{M$_{\odot}$}

\definecolor{smalt(darkpowderblue)}{rgb}{0.0, 0.2, 0.6}
\definecolor{forestgreen(traditional)}{rgb}{0.0, 0.5, 0.0}

\newcommand{\bse}{{\sc bse}}
\newcommand{\mocca}{{\sc mocca}}

\newcommand{\muse}{MUSE}

\newcommand{\chandra}{\emph{Chandra}}

\usepackage{etoolbox}
\makeatletter
\patchcmd\@combinedblfloats{\box\@outputbox}{\unvbox\@outputbox}{}{%
   \errmessage{\noexpand\@combinedblfloats could not be patched}%
}%
 \makeatother

\title[On the absence of SySts in GCs]
{On the absence of symbiotic stars in globular clusters}

\author[D. Belloni et al.]{
\hspace{-0.255cm}
Diogo Belloni$^{1}$\thanks{diogo.belloni@inpe.br (DB)},
Joanna Miko{\l}ajewska$^{2}$\thanks{mikolaj@camk.edu.pl (JM)},
Krystian I{\l}kiewicz$^{2,3}$\thanks{ilkiewicz@camk.edu.pl (KI)},
\newauthor
Matthias R. Schreiber$^{4,5}$,
Mirek Giersz$^{2}$,
Liliana E. Rivera Sandoval$^{3}$
\newauthor
and
Claudia V. Rodrigues$^{1}$
\\
$^{1}$National Institute for Space Research, Av. dos Astronautas, 1758, 12227-010, S\~ao Jos\'e dos Campos, SP, Brazil\\
$^{2}$ Nicolaus Copernicus Astronomical Center, Polish Academy of Sciences, Bartycka 18, 00716 Warsaw, Poland \\
$^{3}$Department of Physics and Astronomy, Box 41051, Science Building, Texas Tech University, Lubbock, TX 79409-1051, USA \\
$^{4}$ Instituto de F{\'i}sica y Astronom{\'i}a, Universidad de Valpara{\'i}so, Av. Gran Breta{\~n}a 1111, 2360102, Valpara{\'i}so, Chile \\
$^{5}$ Millenium Nucleus for Planet Formation, Universidad de Valpara{\'i}so, Av. Gran Breta{\~n}a 1111, 2360102, Valpara{\'i}so, Chile
}

\date{ Accepted 2020 June 11. Received 2020 June 2; in original form 2020 March 17}

\pubyear{2019}

\begin{document}
\label{firstpage}
\pagerange{\pageref{firstpage}--\pageref{lastpage}}
\maketitle

\begin{abstract}
Even though plenty of symbiotic stars (SySts) have been found in the Galactic field and nearby galaxies, not a single one has ever been confirmed in a Galactic globular cluster (GC). We investigate the lack of such systems in GCs for the first time by analysing 144 GC models evolved with the \mocca~code, which have different initial properties and are roughly representative of the Galactic GC population. We focus here on SySts formed through the wind-accretion channel, which can be consistently modelled in binary population synthesis codes. We found that the orbital periods of the majority of such SySts are sufficiently long (${\gtrsim10^3}$~d) so that, for very dense GC models, dynamical interactions play an important role in destroying their progenitors before the present day (${\sim11-12}$~Gyr). In less dense GC models, some SySts are still predicted to exist. However, these systems tend to be located far from the central parts (${\gtrsim70}$~per~cent are far beyond the half-light radius) and are sufficiently rare (${\lesssim1}$~per~GC~per~Myr), which makes their identification rather difficult in observational campaigns. We propose that future searches for SySts in GCs should be performed in the outskirts of nearby low-density GCs with sufficiently long half-mass relaxation times and relatively large Galactocentric distances. Finally, we obtained spectra of the candidate proposed in $\omega$~Cen (SOPS~IV~e-94) and showed that this object is most likely not a SySt.
\end{abstract}


\section{Introduction}
\label{introduction}

Symbiotic Stars (SySts) are
interacting binaries in which,
usually, a white dwarf (WD) accretes matter 
from an evolved red giant 
\citep[see][for a review]
{Mikolajewska_2012}.
They are characterized by
high accretion rates
(greater than a few 
$10^{-9}$~\Msun~yr$^{-1}$),
which are needed to detect the WD 
beside an evolved red giant donor
\citep[e.g.][]{Kenyon_1986},
and sufficiently long orbital periods,
which are needed to accommodate the 
evolved giant.
In many SySts, the accretion rate is high 
enough (greater than a few 
$10^{-8}$~\Msun~yr$^{-1}$)
to trigger and support quasi-steady 
thermonuclear burning. 
The composition of SySts makes them
very important luminous tracers of 
the late phases of low- and medium-mass 
binary star evolution and, in turn, 
excellent laboratories to test models 
of binary evolution.
In particular, 
their studies have important implications 
for, e.g.
understanding mass transfer in wide binaries, 
the interaction of novae with their 
interstellar surroundings, 
or the
formation of astrophysical jets.
Last but not least, 
SySts are also promising nurseries for
type Ia supernovae, regardless 
of whether the path to the thermonuclear 
explosion of a 
Chandrasekhar-mass carbon-oxygen WD is
through accretion, the so-called
single degenerate scenario,
or
through coalescence of double WD systems,
the so-called double degenerate scenario
\citep[e.g.][]{DiStefano_2010,Mikolajewska_2013, Ilkiewicz_2019}.

As in the case of other WD binaries,
such as 
cataclysmic variables
and 
AM CVn,
SySts are usually defined by spectroscopic 
properties
\citep{Kenyon_1986},
i.e.
(i) 
a red continuum with absorption features 
of a late-type red giant;
(ii)
a blue continuum with bright strong 
\mbox{H\,{\sc i}} and \mbox{He\,{\sc i}}
emission lines;
(iii)
either 
additional lines with an ionizational
potential of at least 30 eV 
(e.g. 
\mbox{He\,{\sc ii}},
\mbox{[O\,\sc{ iii}]}, 
\mbox{[Ne\,\sc{ v}]}, and
\mbox{[Fe\,\sc{ vii}]}) 
with
an equivalent width of at least $1$~\AA~
or 
an A- or F-type continuum with additional
absorption lines from 
\mbox{H\,{\sc i}} and \mbox{He\,{\sc i}} 
and singly ionized metals.
This definition seems quite convenient for Galactic SySts, since there is no contamination from the interstellar medium emission lines 
\citep{Mikolajewska_2017}.

Depending on the red giant nature, 
SySts are divided into two main 
classes.
The S-type SySts host normal red giants 
and 
have orbital periods of the order 
of a few years.
The D-type SySts harbour Mira variables
\citep[e.g.][]{Gromadzki_2009b}
usually surrounded by a 
warm dust shell and are expected
to have 
orbital periods of decades or longer
\citep{Whitelock_1987},
despite only one such system 
having a
determined orbital period
\citep[R Aquarii: 43.6~yr,][]
{Gromadzki_2009}.
Even though S-type SySts correspond 
to the majority of known systems
($\sim80$ per cent), the pathways
leading to their formation is far
from being understood, since their 
orbital period distribution 
cannot be accounted for by current
binary populations models
\citep[e.g.][]
{Webbink_1988,Mikolajewska_2012}.

Up to now,
the most detailed study 
of SySts using binary population 
modelling was performed by
\citet{Lu_2006}.
These authors predicted that the 
orbital period distribution of
SySts should peak at 
$\sim1500$~d
and that only 
$\sim20$~per~cent 
SySts should have orbital periods 
shorter than
$\sim1000$~d.
\citet{Lu_2006} explained the discrepancy 
between their result and the observed
orbital period distribution, 
which peaks around $600$~d, 
by an observational incompleteness 
of the sample.
These authors argued that the observations 
were biased towards bright SySts with 
small orbital periods.
At that time, only 30 SySts had
known orbital periods 
\citep{Mikolajewska_2003}.
That sample included
SySts with orbital periods shorter 
than $\sim200$~d, 
which were hardly predicted to exist
in their binary population models.
However, 
since then, 
the orbital periods 
of over 100 known SySts 
in the 
Milky Way 
and  
Magellanic Clouds
have been measured and the main 
characteristics of their 
distribution remain 
practically unchanged 
\citep{Mikolajewska_2012,Gromadzki_2013}.

At the moment, 
there seems to be
a general problem 
with binary population models that 
predict a bi-modal final orbital 
period distribution for binaries 
that have evolved off 
the first giant branch 
and 
the asymptotic giant branch,
in which the common-envelope channel 
results in a rich variety of 
short-period ($\sim1$~day) binaries 
and 
the wind-accretion channel 
results in plenty of systems 
with orbital periods longer 
than $\sim1000$~d 
\citep[][see their fig.~13]{Nie_2012}.
The most peculiar result of the 
adopted evolutionary scenario is 
that there are virtually no binaries 
predicted with orbital periods of 
$\sim100-1000$~d, 
especially because we know they do
exist from observations of both 
SySts and Galactic post-AGB binaries
\citep[e.g.,][]{VanWinckel_2009,Oomen_2018}. 
All these accentuate the need 
for more advanced models for mass 
transfer in binaries with 
red giant donors
\citep[e.g.,][]{Podsiadlowski_2007,Chen_2010,Ilkiewicz_2019}.

SySts have been found since the beginning
of the last century in several different
environments, and more than 200
such systems exist in the Milky Way
\citep[e.g.,][]
{Belczynski_2000,Miszalski_2013,Miszalski_2014,R_Flores_2014,Merc_2019}.
In addition, there are plenty discovered
in nearby galaxies, such as 
the Magellanic Clouds
\citep{Ilkiewicz_2018},
M31
\citep{Mikolajewska_2014},
M33
\citep{Mikolajewska_2017} 
as well as single SySts in 
NGC~6822 
\citep{Kniazev_2009} 
and 
NGC~205
\citep{Goncalves_2015}.
Despite the frequency of SySts in several
different environments, not a single
one has ever been detected in a
Galactic globular cluster (GC).
So far, only a few attempts have been made 
to identify SySts in Galactic GCs.
However, such investigations have
not been designed for that purpose
and, as pointed out by 
\citet{Zurek_2016},
due to their long orbital periods 
and 
the dominant contribution of the red giant
at longer wavelengths, 
SySts will usually be missed
by optical variability surveys.
Spectroscopic surveys are ideal to 
identify emission lines and, in turn,
SySts,
but they are very time-consuming 
and, consequently, rare. 
However,
photometric surveys using narrow-band 
filters centred on 
either
the \mbox{He\,{\sc ii}} and H$\alpha$ 
\citep{Ilkiewicz_2018} 
or 
the Raman-scattered 
\mbox{[O\,\sc{ vi}]}
emission lines 
\citep{Angeloni_2019}
are another promising way to look 
for SySts in GCs.

The first SySt thought to be related to a GC is Pt~1, possibly associated with the GC NGC~6401 \citep{Peterson_1977}. This system though was later classified as a Galactic halo SySt \citep{TorresPeimbert_1980}.
After that, \citet{Zurek_2016} suggested that the far-ultraviolet variable source N1851-FUV1, within the core of NGC~1851, could be a SySt, given that there is a red giant spatially coincident with this source.
However, its spectrum clearly lacks any emission lines, which indicates that the SySt interpretation is probably not right and the presence of a red giant nearby is just a chance superposition of two unrelated objects.
This source is now believed to be an AM CVn candidate, based on its X-ray properties, its spectral energy distribution and the amplitude of its light curve.
More recently, \citet{Henleywillis_2018} proposed that the second-brightest X-ray source in $\omega$~Cen, possibly associated with the carbon star SOPS~IV~e-94,
is a promising SySt candidate.
As we shall see in Section~\ref{SyStOmega}, based on published and new data, SOPS~IV~e-94 is most likely not a SySt.

GCs are one of the most important objects 
for investigating the formation and the 
physical nature of exotic systems
such as 
X-ray binaries, 
degenerate binaries, 
black holes,
blue straggler stars, 
cataclysmic variables,
millisecond pulsars,
etc
\citep[e.g.][]{Benacquista_2013}.
Such studies provide
tools that can help to understand 
the formation and evolution processes 
of star clusters, galaxies and, 
in general, the young Universe. 
Therefore, understanding the absence
of SySts in GCs might lead to important
astrophysical implications.

We concentrate here on SySts formed
through the wind-accretion channel, i.e.
without Roche-lobe overflow in the WD
formation.
We notice that most SySts are S-type 
and, in most of them, the WD likely 
formed in an episode of 
Roche-lobe overflow.
However, their formation channels 
are clearly not understood, which 
makes the modelling of these systems 
difficult, not only in isolation,
but also in GCs.
Therefore, we leave these systems 
for follow-up works, in which we will 
first try to explain their 
orbital period distribution, 
and subsequently investigate the 
role of dynamics in shaping their 
properties in GCs.

In this paper,
we search for the physical reasons
behind the absence of SySts in GCs.
In particular, we check whether
dynamics could play a significantly important
role in destroying their 
progenitors during the GC evolution.
In addition, for those GC SySt
that are not destroyed,
we predict their properties
and correlations with their host GCs,
by providing relevant information that might
help future theoretical and observational 
efforts.

\section{Is SOPS~IV~e-94 a symbiotic star in $\omega$~Cen?}
\label{SyStOmega}

The second-brightest X-ray source in $\omega$~Cen, {CXOHCD~J132601.59--473305.8}, lies at about $8.8$~arcmin southwest of the cluster centre.
The position of this X-ray source coincides closely with that of SOPS~IV~e-94, which is a Population~II carbon star \citep[][]{Harding_1962} and the first such a star identified in a GC.
Indeed, SOPS~IV~e-94 is at $\sim0.34$~arcsec from the \chandra~position of {CXOHCD~J132601.59--473305.8}, inside the $95$~per~cent confidence radius of $\sim0.55$~arcsec. 
\citet{vanLoon_2007} noticed that it is the brightest and reddest carbon star in the cluster, and its very high 12C:13C ratio points at the $s$-process in an asymptotic giant branch carbon star to have been responsible for its large carbon overabundance.
Based on the characteristics of carbon stars and the optical and X-ray properties of this source, \citet{Henleywillis_2018} proposed that this could be the first SySt ever identified in a Galactic GC.

\begin{figure}
\begin{center}
\includegraphics[width=0.99\linewidth]
    {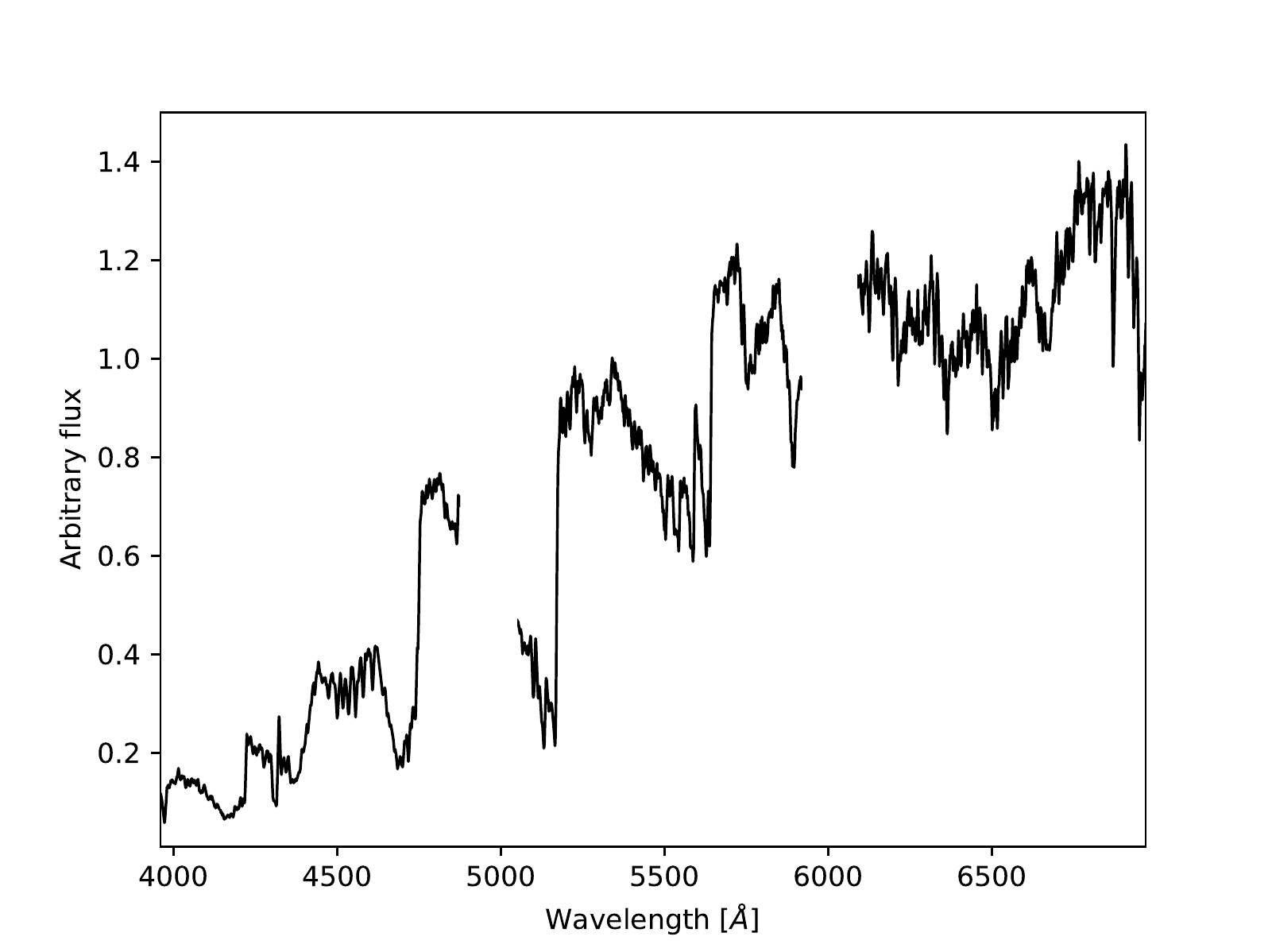}
\includegraphics[width=0.99\linewidth]
    {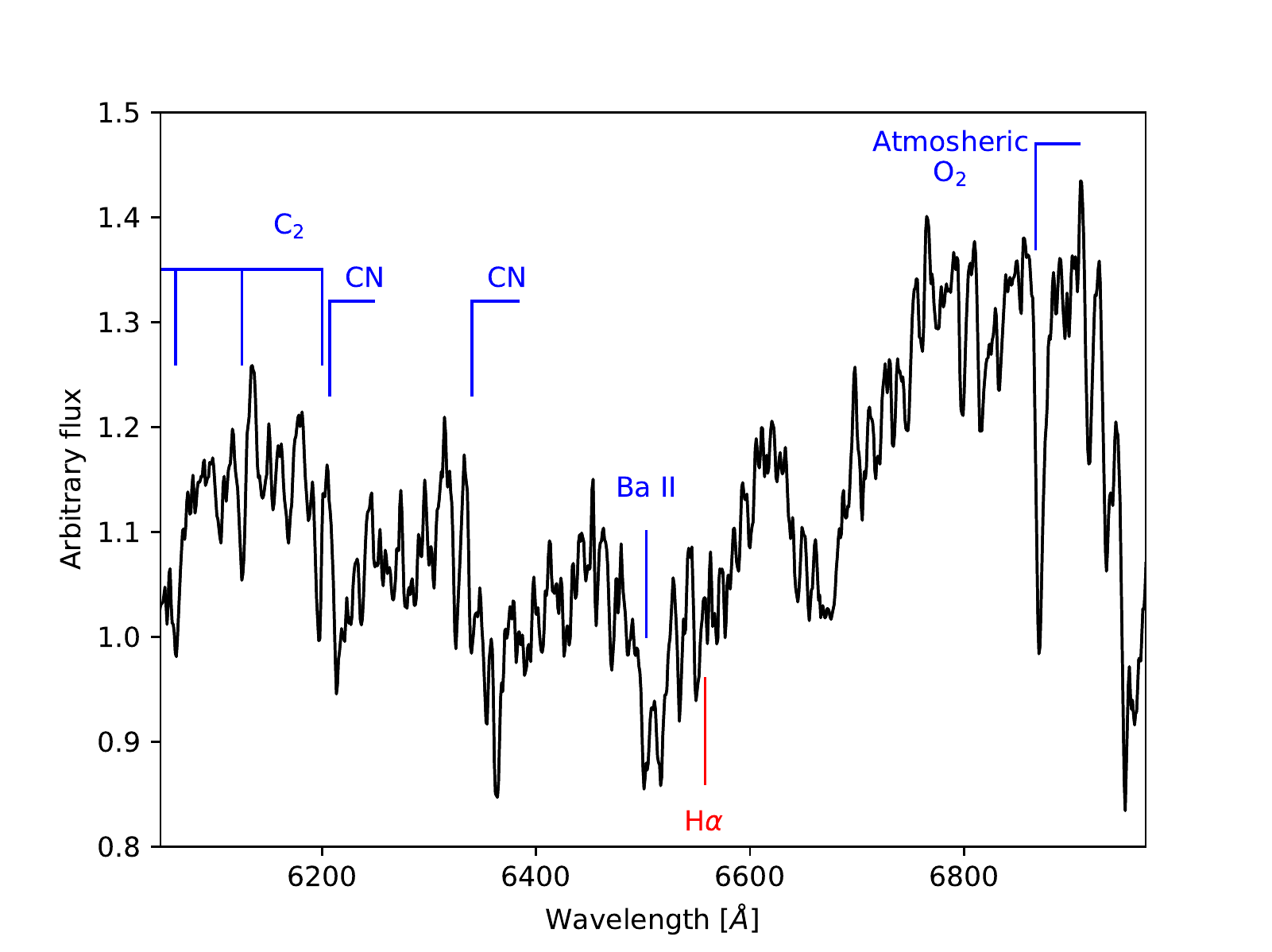}
\end{center}
\caption{SALT spectrum of 
SOPS~IV~e-94. 
\textit{Top panel}: full spectrum. 
\textit{Bottom panel}: region of the 
spectrum near the expected H$\alpha$ line, 
with the main spectral features identified 
and a visible lack of H$\alpha$ line.
}
  \label{FigSpectrum}
\end{figure}

In order to identify SOPS~IV~e-94 as a SySt, 
particular features in the spectrum are 
expected (see Sec.~\ref{introduction}) 
and until these are indeed observed, 
this object cannot be confidently 
classified as a SySt.
The minimum requirement is the presence 
of  Balmer emission lines
in the optical spectrum.
Whereas the published optical photometry 
of {SOPS~IV~e-94} 
\citep
[${U=14.467}$, ${B=13.301}$, ${V=11.517}$, ${R_{\rm C}=10.691}$, ${I_{\rm C}=9.917}$, ${{\rm H}\alpha=10.263}$;][]
{Bellini_2009}
indicates that there could be some 
H$\alpha$ emission 
(${{\rm H}\alpha-R_{\rm C}=-0.428}$), 
the published spectra 
\citep[][]{Harding_1962,vanLoon_2007}
do not show any emission lines.
Instead, 
the negative H$\alpha$ $-$ $R_{\rm C}$ 
colour could be simply due to strong 
molecular bands that are present in 
the optical spectrum and affect the 
$R_{\rm C}$ mag.
In such a case, 
SOPS~IV~e-94 would represent another 
example of a chance superposition of 
two unrelated objects -- the X-ray 
source and the carbon giant -- as 
in the case of N1851-FUV1 in 
NGC~1851.

To refine the nature of SOPS~IV~e-94,
we obtained a deep optical spectrum 
with the Robert Stobie Spectrograph 
\citep[RSS;][]{Burgh_2003N,Kobulnicky_2003N}
mounted on the 
Southern African Large Telescope 
\citep[SALT;][]{Buckley_2006N,Donoghue_2006} 
under programme 2019-2-SCI-021 
(PI: I\l{}kiewicz). 
A single RSS configuration was adopted 
with PG900 grating and a slit width of 
1.5~arcsec to give wavelength coverage from 
$\sim3920$ to $\sim6980$~\AA~ with resolving power
${\rm R}\simeq1000$.
The spectrum, presented in 
Fig.~\ref{FigSpectrum}, was made on 
14-01-2020 with a 
$200$~s exposure time.

The spectrum of SOPS~IV~e-94 is 
dominated by strong C$_2$ and CN 
molecular bands typical for 
a carbon star.
At the same time, 
none of the features typical for 
a SySt are present.
In particular,
there is no trace of any 
emission lines, including the 
strongest H$\alpha$ emission 
which is always visible in SySts.
Given the X-ray luminosity of 
{CXOHCD~J132601.59--473305.8}, 
which is comparable to known SySts 
\citep[][and references therein]{Henleywillis_2018},
one should expect to detect
at least the 
brightest \mbox{H\,{\sc i}} Balmer 
emission lines.
The lack of any emission lines in the 
spectrum of the carbon star indicates 
that there is no physical association 
between this star and 
{CXOHCD~J132601.59--473305.8},
and the X-ray emission most probably 
originates from another object.
Therefore, 
{\it
SOPS~IV~e-94 cannot 
be classified as a SySt.
}

\section{Numerical Simulations}
\label{approach}

In what follows, 
we briefly describe the 
GC models and the \mocca~code 
\citep[][and references therein]
{Hypki_2013,Giersz_2013} used to
simulate them.  
More details 
about the modelling/models can be 
found in \citet{Belloni_2019}.

\subsection{Globular Clusters}

\mocca~includes the {\sc fewbody} 
code \citep{Fregeau_2004} to perform 
numerical scattering experiments of 
small-number gravitational interactions 
and the \bse~code 
\citep{Hurley_2000,Hurley_2002}, with the 
upgrades described in \citet{Belloni_2018b}  
and \citet{Giacobbo_2018}, to deal with 
stellar and binary evolution.
This version of the \mocca~code includes 
up-to-date prescriptions for 
metallicity-dependent stellar winds,
which are based on
\citet{Belczynski_2010},
but with the inclusion of the
Eddington factor
from
\citet{Chen_2015}.

\mocca~assumes a point-mass 
Galactic potential with total 
mass equal to the enclosed Galaxy 
mass inside a circular orbit at 
the specified Galactocentric radius, 
and uses the description of escape processes 
in tidally limited clusters follows the 
procedure derived by 
\citet{Fukushige_2000}. 
We stress that \mocca~has been extensively 
tested against $N$-body codes and reproduces 
$N$-body results with good precision,
including detailed distributions of mass 
and binding energy of binaries
\citep[e.g.][]
{Giersz_2008,Giersz_2013,Wang_2016,Madrid_2017}. 
Most importantly, 
\mocca~is faster than $N$-body codes, 
which allows us to simulate several 
hundreds of real GC models that permit 
more powerful statistical analyses 
for constraining the overall population 
of particular types of binaries in GCs.

In all models, we assume that all stars
are on the zero-age main sequence
when the simulation begins and that any 
residual gas from the star
formation process has already been removed 
from the cluster.
Additionally,
all models have low metallicity ($Z=0.001$), 
are initially at virial equilibrium, and
have neither rotation nor mass segregation. 
With respect to the density profile, 
all models follow a \citet{King_1966} model,
and we adopted two values for the King 
parameter W$_0$: 6 and 9.
Regarding the tidal radius, we assumed 
two values, namely 60 and 120 pc.
Finally, we have three different 
half-mass radii: 1.2, 2.4 and 4.8 pc. 

The initial binary population adopted 
here for all models corresponds to models 
constructed based on the distributions
derived by 
\citet{Kroupa_1995b,Kroupa_INITIAL}
and
\citet{Kroupa_2013}, 
with the modifications described in 
\citet{Belloni_2017c}.
We simulated models with three different 
numbers of objects (single stars $+$ binaries),
namely $400$k, $700$k, and $1200$k,
which have masses of approximately 
$4.72 \times 10^5$, 
$8.26 \times 10^5$ 
and 
$1.42 \times 10^6$  ${\rm M_\odot}$,
respectively.
All of them have very high initial
binary fraction
\citep[nearly 100~per~cent, e.g.][]{Kroupa_INITIAL},
which is needed to resolve the angular 
momentum problem in star formation 
and 
consistent with the fact that triples 
and higher order systems are rarely 
the outcome of star formation 
\citep[e.g.][]{Goodwin_2005}.
In all models, we have used the 
\citet{Kroupa_2001} canonical initial
mass function, with
star masses in the range between 
$0.08~{\rm M_\odot}$ and $150~{\rm M_\odot}$
\citep[][]{Weidner_2013}.

For each initial cluster configuration, 
we simulated models with three values
for the common-envelope efficiency, 
namely 
$0.25$, $0.5$ and $1.0$.
In addition, we assumed that none of the
recombination energy helps in the 
common-envelope ejection 
and that the binding energy parameter is 
automatically determined based on the giant
properties
\citep[][appendix A]{Claeys_2014}.
Even though we focus on SySts in which
the WDs are formed without Roche-lobe 
overflow, several other types of binaries 
inside the cluster are affected by the
choice of the CE efficiency. 
This choice thus influences the amount 
of particular types of GC binaries 
\citep[see][for the case of cataclysmic variables]{Belloni_2019}.
In this way,
even though this has never
been thoroughly checked,
the choice of the CE efficiency 
may play a role in the global
GC evolution.

For massive stars, we assumed
the delayed core-collapse supernova model
\citep{Fryer_2012}. 
We also included
pair-instability supernovae 
and 
pair-instability pulsation supernovae,
as described in \citet{Spera_2017}.
Supernova natal kicks 
for neutron stars are distributed according 
to the Maxwellian distribution suggested by 
\citet{Hobbs_2005}.
In the case of black holes, we have two options: 
either kicks are distributed according to 
\citet{Hobbs_2005} and no fallback prescription 
is adopted;
or kicks follow \citet{Hobbs_2005} and are 
reduced according to mass fallback description 
given by \citet{Fryer_2012},
for the delayed core-collapse model.
As part of the upgrades to the \bse~code, 
we included in our modelling the possibility 
of neutron star formation through 
electron-capture supernova
\citep[e.g.][]{Kiel_2008}
and accretion induced collapse
\citep[e.g.][]{Michel_1987}.
In both cases, we assume no kick associated 
with the neutron star formation.
All other binary evolution parameters are 
set as in \citet{Hurley_2002}.

\begin{table} 
\caption{Initial GC conditions and 
binary evolution parameters.
For all models, 
we adopted a  low metallicity ($Z=0.001$),
the canonical \citet{Kroupa_2001} 
initial mass function,
with masses between 
0.08 M$_\odot$ and 150 M$_\odot$,
and
a high binary fraction (95~per~cent).}
\label{TabMODELS}
\begin{adjustbox}{width=210px}
\noindent
\begin{threeparttable}
\noindent
\begin{tabular}{l|c|l}
\hline\hline
Property	& & Values		\\
\hline
Number of objects ($\times10^5$)	& & 4, 7, 12		\\
Mass ($\times10^5$~M$_\odot$)		& & 4.72, 8.26, 14.2 	\\
King model parameter				& & 6, 9				\\
Tidal radius (pc)					& & 60, 120			\\
Half-mass radius (pc)				& & 1.2, 2.4, 4.8		\\
Fallback						    & & yes, no			\\
Common-envelope efficiency 			& & 0.25, 0.50, 1.00 		\\
\hline\hline
\end {tabular}
\end{threeparttable}
\end{adjustbox}
\end{table}

All the parameters and initial GC conditions
discussed above 
are summarized in 
Table~\ref{TabMODELS}.
As shown by \citet{Belloni_2019},
by comparing the 
simulated and observed
distributions of 
core to half-light radii,
$V$-band absolute magnitude,
average surface brightness inside 
the half-light radius and
central surface brightness,
our models are very close to
massive and intermediate-mass
real GCs, and we only miss the 
low-mass GCs in our analysis.
Additionally, our present-day GC models
cover a reasonable range of 
concentrations, 
central surface brightness 
and 
half-mass relaxation times
\citep[see also][]{Askar_2016b}.
Therefore, our models are consistent 
with a substantial fraction of the real GCs,
and are in turn roughly representative 
of the Galactic GC population.

\subsection{Symbiotic Stars}

Our principal goal here is to investigate the 
properties of SySts formed through the 
wind-accretion channel in our simulations.
For that end, 
we define SySts as 
WD\,+\,red giant binaries, 
in which the WDs are formed
avoiding Roche-lobe overflow.
In addition,
those binaries are SySts only if
their accretion-powered WD luminosities
are at least 10~\Lsun~
\citep[e.g.][]
{Mikolajewska_1992,Lu_2006}.
This is the luminosity resulting from
the release of gravitational 
energy due to accretion and
is given by

\begin{equation}
\frac{L_{\rm WD}}{{\rm L}_\odot} \ \approx \ 
3.14\times10^7~
     \left( 
\frac{M_{\rm WD}}{{\rm M}_\odot} 
     \right)
     \left( 
\frac{\dot{M}_{\rm WD}}{{\rm M}_\odot~{\rm yr}^{-1}} 
     \right)
     \left( 
\frac{R_{\rm WD}}{{\rm R}_\odot} 
     \right)^{-1},
\label{Lwdgrav}
\end{equation}
\

\noindent
where
$M_{\rm WD}$ is the WD mass,
$R_{\rm WD}$ is the WD radius
and
$\dot{M}_{\rm WD}$ is the accretion
rate onto the WD.
Given such high accretion luminosities,
red giants in SySts are usually located
towards the top of 
either the first giant branch 
or the asymptotic giant branch
\citep[e.g.][]{Mikolajewska_2007}.

In the \bse~code, the accretion rate
efficiency of mass loss through winds
is estimated according to the 
\citet{BH44}
mechanism, given by
\begin{equation}
\beta_{\rm BH} 
= 
\frac{\alpha_{\rm BH}}{2\sqrt{1-e^2}}
    \left(
      \frac{GM_{\rm WD}}{a~v_\mathrm{w}^2}
    \right)^2~\left[1 + \left(\frac{v_{\mathrm{orb}}}{v_{\mathrm{w}}}\right)^2\right]^{-3/2}
    ~,~\label{BH}
\end{equation} 
\

\noindent
where 
$G$ is the gravitational constant,
$v_\mathrm{w}$ and $v_\mathrm{orb}$ 
are the wind and orbital velocities,
respectively,
$a$ is the semi-major axis,
$e$ is the eccentricity, 
and 
$\alpha_{\rm BH}=1.5$.

This prescription is known to 
underestimate the efficiency of 
wind mass transfer in binaries, 
especially in the case of red giants
in the asymptotic giant branch,
which have slow and dense winds.
Thus, to properly identify the 
SySts in our simulations, 
we implemented into the \bse~code 
the wind Roche-lobe overflow
mechanism,
as described in 
\citet{Abate_2013}
and
\citet{Ilkiewicz_2019}.
Briefly,
the enhanced accretion efficiency is
given by
\citet{Abate_2013}:

\begin{equation}
\beta_{\rm WRLOF} \ = \
\frac{25}{9}\,q^2 
\left[
-0.284
  \left(
\frac{R_d}{R_{\rm RL}}
  \right)^2
+0.918
  \frac{R_d}{R_{\rm RL}}-0.234 
\right]
~,~
\label{WRLOF}
\end{equation}
\

\noindent
where 
$q=M_{\rm WD}/M_{\rm giant}$,
$M_{\rm giant}$ is the red giant mass,
$R_{\rm RL}$ is the red giant Roche-lobe radius
and
$R_d$ is the dust condensation radius 
given by
\citet{Hofner_2007}:

\begin{equation}
R_d
=
\left( 
\frac{R_{\rm giant}}{2}
\right)
\left( 
\frac{T_d}{T_{\rm eff,giant}}
\right)^{-(4+p)/2},
\end{equation}
\

\noindent
where 
$R_{\rm giant}$ is the giant radius,
$T_{\rm eff,giant}$ is the giant effective temperature,
$T_d$  is the dust condensation temperature, 
and 
$p$ is a parameter characterising 
wavelength dependence of the dust 
opacity.

The WD cannot accrete more 
mass than is lost by the red giant, 
which might happen for highly 
eccentric systems with Eq.~\ref{BH}.
To avoid this,
as in \citet{Hurley_2002},
we enforced that 
{${\beta_{\rm BH}\leq0.8}$}.
In addition, 
as in \citet{Abate_2013},
we imposed that
{${\beta_{\rm WRLOF}\leq0.5}$}
to be consistent with results
from hydro-dynamical simulations. 
Moreover,
we assumed dust consisting of 
amorphous carbon grains, which
gives 
{${T_d\simeq1\,500}$\,K}
and 
{${p\simeq1}$}
\citep{Hofner_2007}.
Finally,
as in \citet{Ilkiewicz_2019},
in our simulations, having calculated 
both 
wind Roche-lobe overflow
(Eq.~\ref{WRLOF}) 
and 
Bondi-Hoyle
(Eq.~\ref{BH})
accretion rate efficiencies, 
we took the higher to be 
the accretion rate efficiency.

We would like to stress that several potentially important parameters are kept constant, such as the metallicity, ${\alpha_{\rm BH}}$, the initial mass function, the initial binary population, amongst others, because simultaneously varying all possible parameters in GC modelling is not feasible. As such parameters could impact some of the results presented here, simulations in which the explored parameter space is extended will be useful to further test the results achieved in this work.

\section{Symbiotic Star Properties}
\label{resultsSYMB}

We start the presentation of
our results by focusing on the
initial and present-day
properties of the simulated
SySts in GCs and
in isolated binary evolution,
i.e. without dynamics.
The 
\textit{initial} 
time corresponds to the
beginning of the simulation, i.e.
roughly when the cluster is born,
while
the 
\textit{present-day} 
time is
assumed here to be $\sim11-12$~Gyr,
which is consistent with the 
measured ages of Galactic GCs
\citep{VandenBerg_2013}.
In order to obtain the properties 
of SySts in a non-crowded 
environment,
we selected all the initial binary
populations 
(composed of zero-age main-sequence binaries),
which follow the same
distributions
\citep{Belloni_2017c}
in all models, and evolved them
with the \bse~code till the 
present day.

\subsection{Orbital Period}

\begin{figure}
\begin{center}
\includegraphics[width=0.975\linewidth]{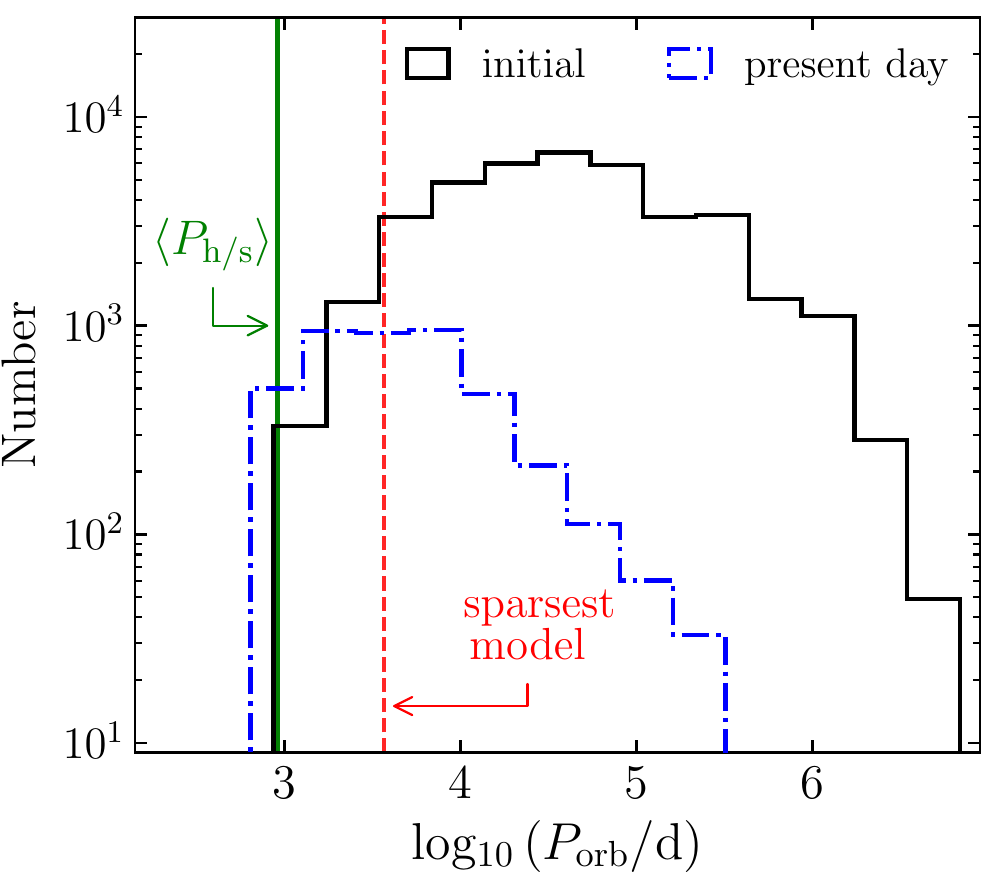} 
\end{center}
\caption{Orbital period distributions of SySt progenitors at the beginning of the simulations (black solid histogram) and of present-day SySts inside the GC models (dot-dashed blue lines). The green solid vertical line is the average initial hard-soft boundary $\langle P_{\rm h/s}\rangle$ in our GC models and the red dashed vertical line corresponds to the $P_{\rm h/s}$ of the sparsest model. For very dense models, with very short $P_{\rm h/s}$, most (if not all) SySt progenitors are expected to be destroyed during the GC evolution, as their orbital periods are considerably much longer than $P_{\rm h/s}$, which makes them rather soft and easily destroyed in dynamical interactions. On the other hand, a substantial fraction of SySts might still survive in the outer parts of less dense clusters, as their orbital periods are comparable to $P_{\rm h/s}$.}
\label{Fig01}
\end{figure}

We show in Fig.~\ref{Fig01} the initial orbital period distribution of SySt progenitors and the orbital period distribution of present-day GC SySts.
The present-day orbital periods are those in which, during the SySt phase, the accretion-powered WD luminosity is highest.
The orbital periods are sufficiently long to put SySts amongst the largest interacting binaries.
Present-day orbital periods range from $\sim10^3$ to $\sim10^5$~d, while initial orbital periods of SySt progenitors extend up to $\sim10^6$.
As one can see, our predicted orbital periods more likely resemble those of D-type SySts than S-type SySts.
Moreover, while comparing both distributions in Fig.~\ref{Fig01}, we can see the role played by dynamics in shaping the parameter space of GC SySts and their progenitors.
Due to dynamics, a lot of progenitors are disrupted, limiting the orbital periods from being very long.

In the same figure, we also show the
average initial 
\textit{hard-soft boundary}
$\langle P_{\rm h/s}\rangle$
in our GC models
($\sim10^3$~d),
and 
that of the sparsest model
($\sim10^{3.6}$~d).
This boundary is set when the
average binary binding energy 
equals to the
average cluster kinetic energy.
This separation is thus
intrinsically related
to the interplay between 
the binary binding energies
with respect to host GC properties.
Pragmatically,
it corresponds to the
orbital period separating 
hard 
$(P_{\rm orb}<P_{\rm h/s})$ 
and 
soft 
$(P_{\rm orb}>P_{\rm h/s})$ 
binaries
\citep[e.g.][]{HeggieBOOK}.
\textit{Hard} binaries are very strongly 
bound and are not expected to go through
disruptive encounters.
\textit{Soft} binaries, on the other hand,
are very weakly bound and tend to be 
destroyed in dynamical interactions.
Some binaries have orbital periods 
comparable to the hard-soft boundary 
and can sometimes be destroyed
or only significantly altered.
Most binaries,
on average,
evolve according to 
the Heggie--Hills law: 
{\it
hard binaries get harder, 
while soft binaries get 
softer, 
after dynamical interactions
}
\citep{Heggie_1975,Hills_1975},
which implies that soft binaries
tend to be eventually disrupted.

The orbital period defining the
hard-soft boundary, based on average
properties, in a particular
GC is given by

\begin{equation}
\frac{P_{\rm h/s}}{{\rm yr}} \ = \ 
\sqrt{
   \bigg(
\frac
{a_{\rm h/s}}{{\rm au}}
   \bigg)^3
   \bigg(
2\,\frac
{\langle m \rangle}
{{\rm M}_\odot}
   \bigg)^{-1}
     },
\label{phs}
\end{equation}
\

\noindent
where 
$\langle m \rangle$ is the average
mass, given by $M_{\rm GC}/N$, where
$N$ is the number of objects 
(single $+$ binaries),
$M_{\rm GC}$ is the total mass,
and
$a_{\rm h/s}$ is 
the semi-major axis that defines the 
hard-soft boundary and is given by
$R_{\rm half{\text -}mass}/0.4N$, 
where $R_{\rm half{\text -}mass}$ 
is the half-mass radius
\citep{Spitzer_BOOK}.
We can safely apply Eq.~\ref{phs}
since SySts are not much more massive 
than an average star/binary in a cluster.
More specifically,
they are probably about 
two to three times more massive 
(red giants about $1.5$ times 
and 
WD about $1.5$ times).
Thus, 
the time-scale for SySts being 
mass segregating is not extremely 
short.
They need more than the half-mass relaxation 
time to sink to the centre from the GC halo 
(farther than the half-mass radius).

For clusters with similar 
$N$, Eq.~\ref{phs} says that
the denser the cluster
(i.e. the smaller the half-mass radius), 
the smaller the semi-major axis 
(or the shorter the orbital period) 
that defines the hard-soft boundary.
Thus, at a particular density,
the hard-soft boundary will penetrate the
region occupied by SySt progenitors, as
illustrated in Fig.~\ref{Fig01}.
Therefore, beyond this density, 
more and more SySt progenitors 
are potentially destroyed, as the 
density increases.
As mentioned before, the fate of 
SySt progenitors
with orbital periods comparable to
the orbital period defining the hard-soft
boundary is not so easy to predict.
Therefore, even though we expect many
of them to be destroyed before the
present-day, some might potentially
survive the GC dynamical evolution.
Indeed, those binaries with the 
shortest periods 
($\lesssim10^4$~d)
in the distribution might
survive in less crowded regions
inside the clusters, as the
probability for interaction
in such regions are much smaller
than in the central parts.
This is especially true for
clusters with sufficiently long
initial half-mass
relaxation times, since in these
clusters mass segregation is not
very efficient.

\subsection{Other Properties}

\begin{figure*}
\begin{center}
%
%
\includegraphics[width=0.33\linewidth]
{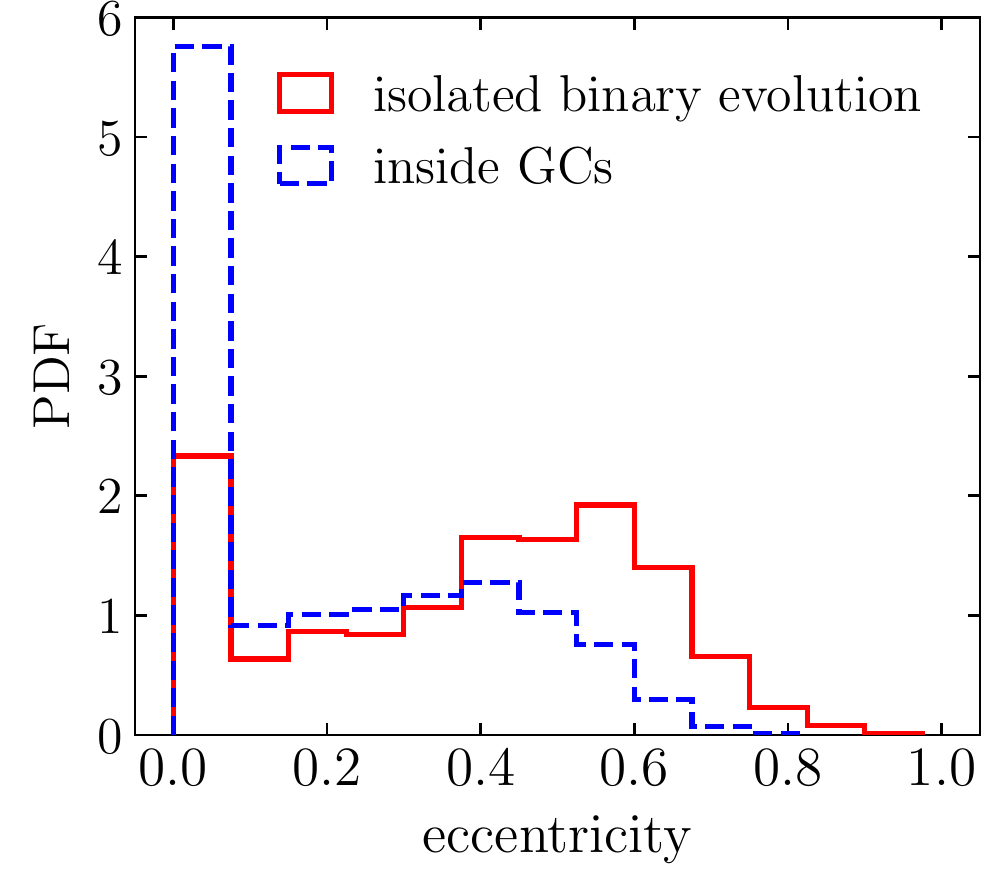}
\includegraphics[width=0.33\linewidth]
{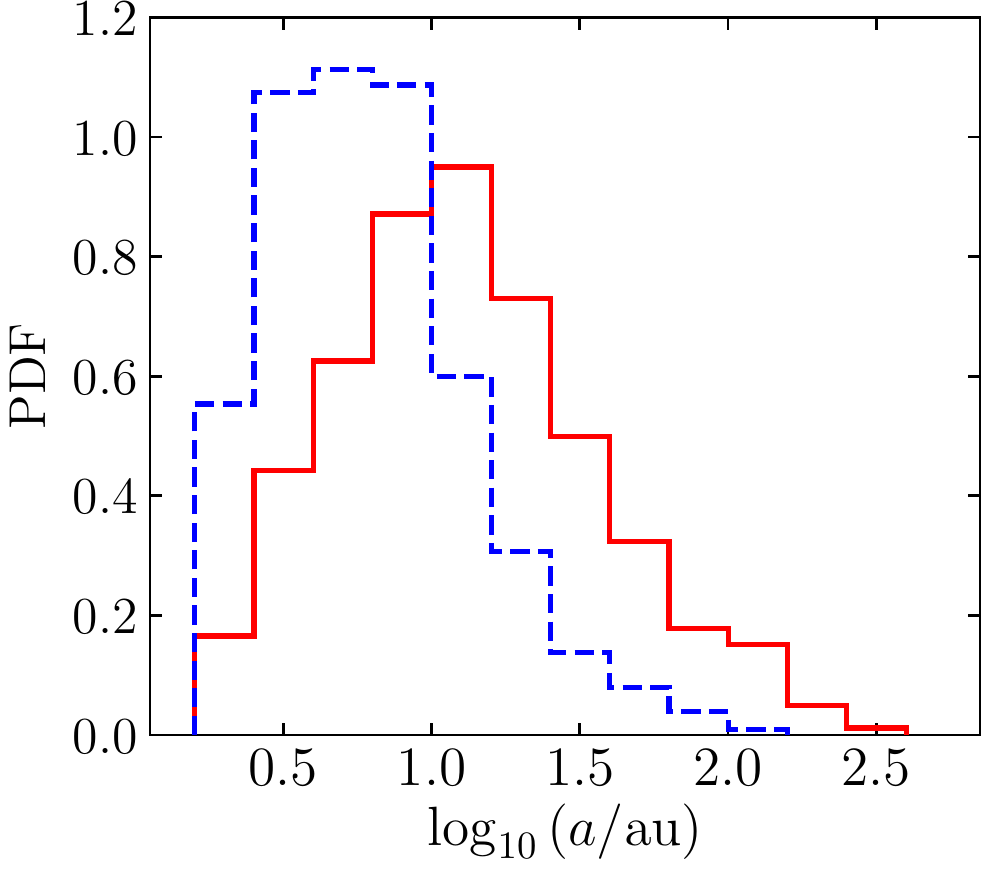}
\includegraphics[width=0.33\linewidth]
{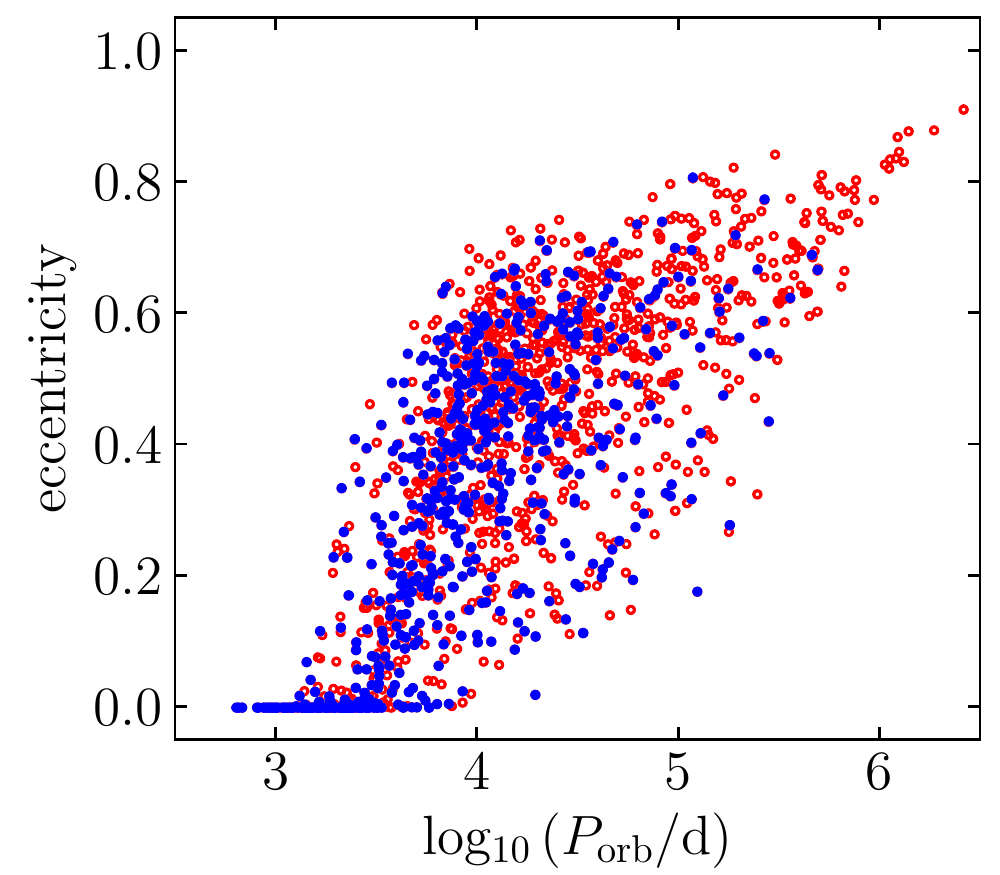}
%
%
\includegraphics[width=0.33\linewidth]
{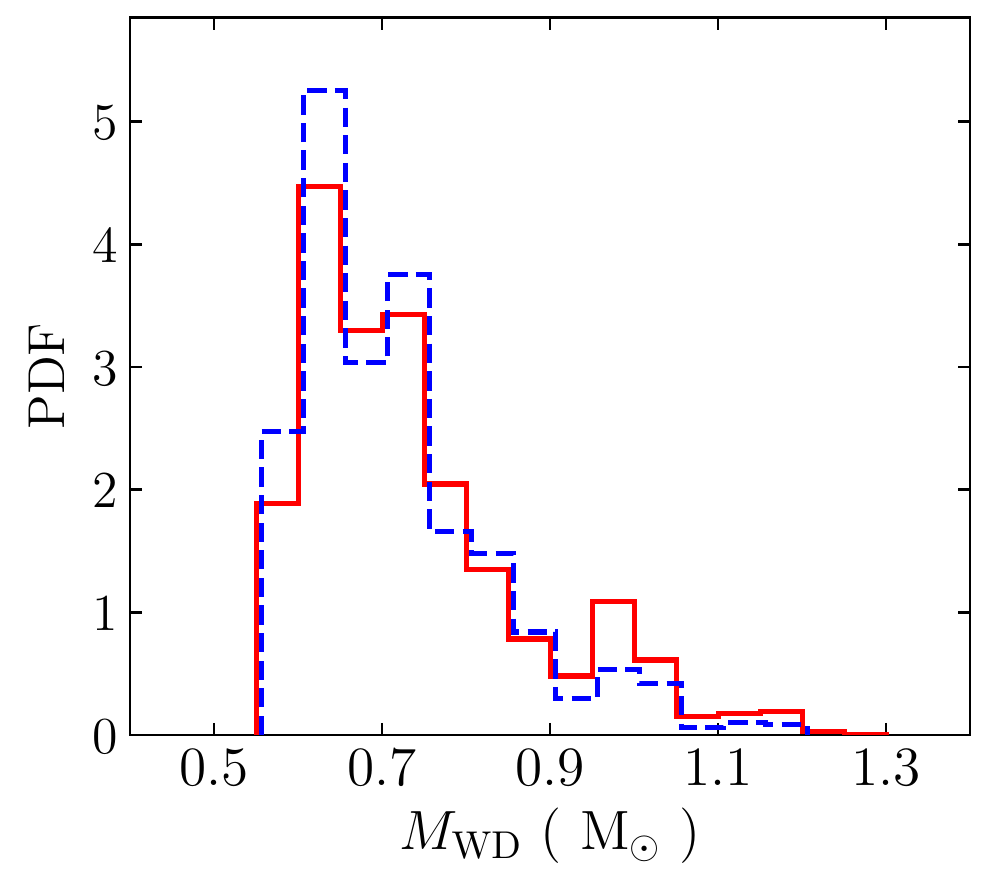}
\includegraphics[width=0.33\linewidth]
{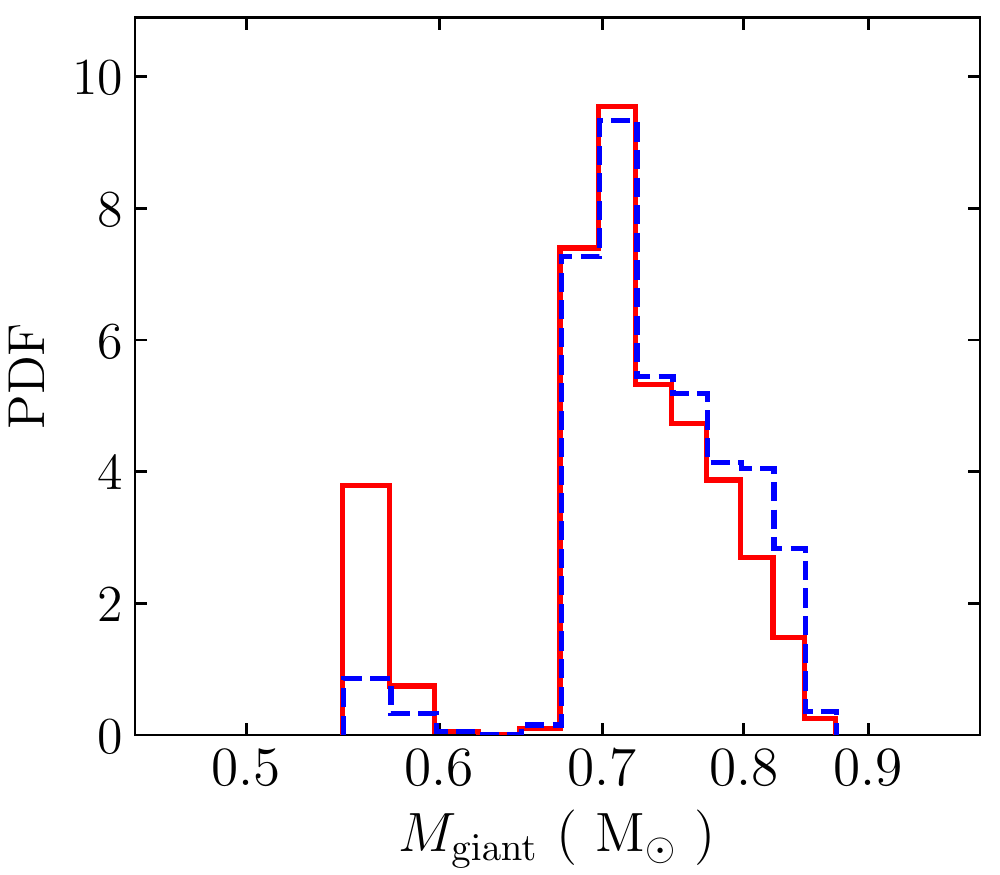}
\includegraphics[width=0.33\linewidth]
{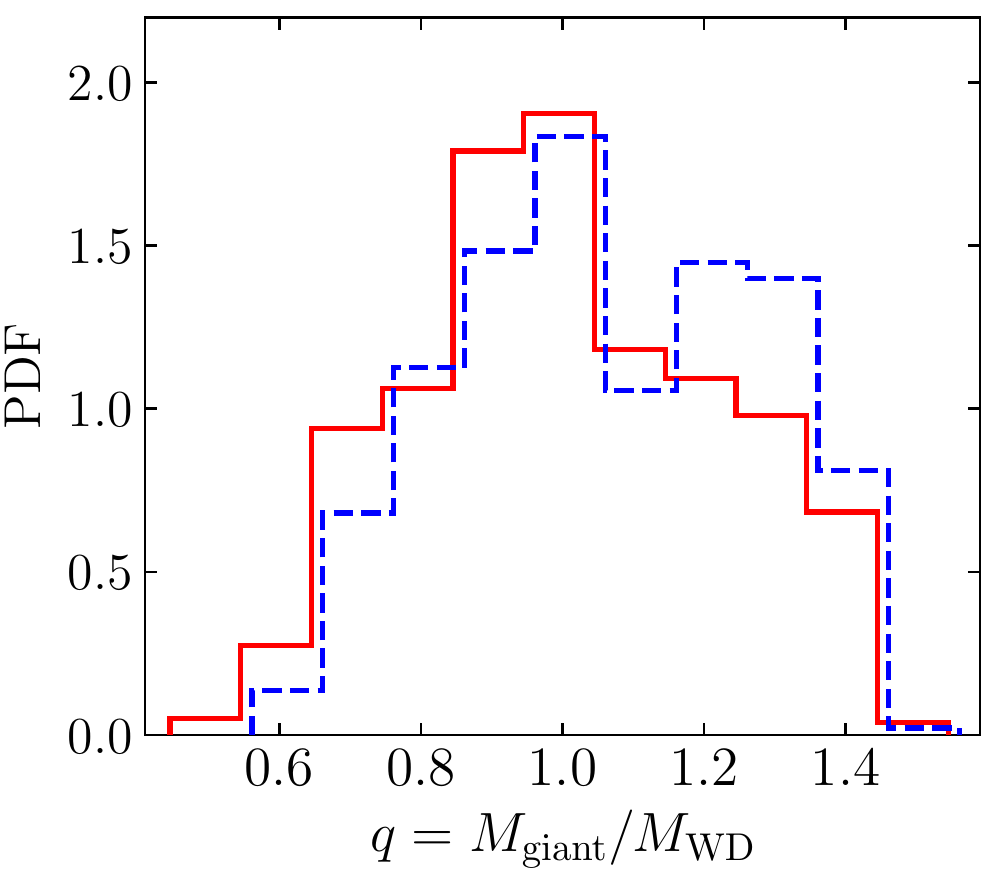}
%
%
\includegraphics[width=0.33\linewidth]
{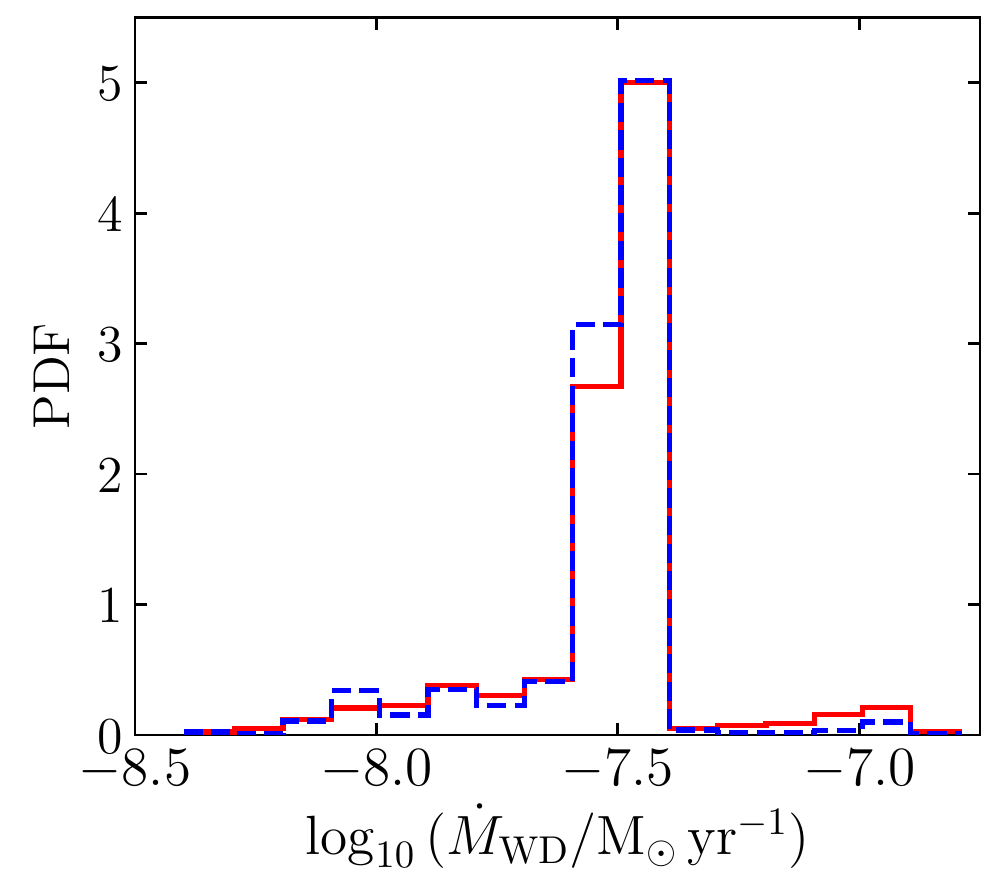}
\includegraphics[width=0.33\linewidth]
{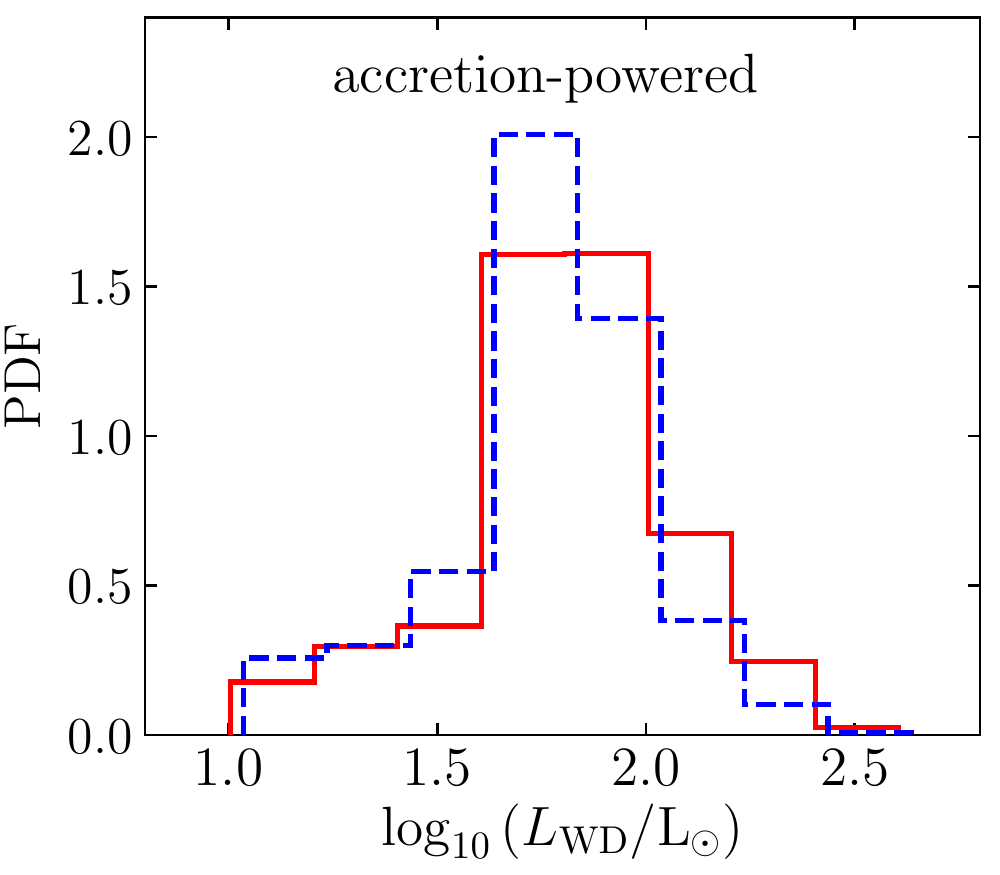}
\includegraphics[width=0.33\linewidth]
{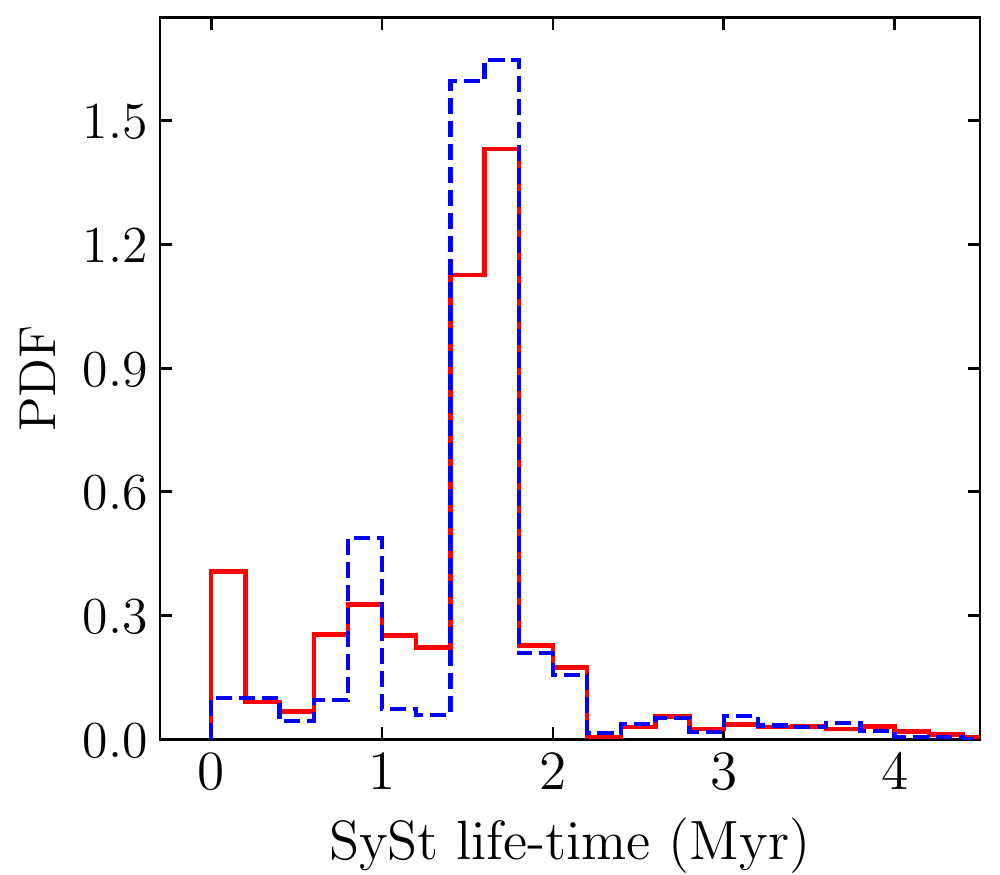}
\end{center}
\caption{Distributions of present-day properties of SySts in GCs
(either dashed blue histograms or filled blue circles)
and obtained in isolated binary evolution,
i.e. without dynamics
(either solid red histograms or open red circles), 
namely
eccentricity 
(top left-hand panel),
semi-major axis 
(top middle panel),
plane eccentricity versus orbital period
(top right-hand panel),
WD mass 
(middle left-hand panel),
red giant mass
(middle middle panel),
mass ratio
(middle right-hand panel),
accretion rate onto the WD 
(bottom left-hand panel),
accretion-powered WD luminosity 
(bottom middle panel),
and
SySt life-time
(bottom right-hand panel).
Distributions are normalized such 
that the area under
each histogram is equal $1$
(probability distribution function, PDF).
Notice that, with the exception of
the 
eccentricity,
semi-major axis
and
orbital period,
distributions are usually
similar, while comparing GC SySts with
those that would be produced in a
non-crowded environment with the same
metallicity and star formation history.
The reduced range in 
the eccentricity,
semi-major axis 
and 
orbital period
distributions of GC SySts 
is due to dynamics,
which play a major role in destroying the
systems with longest orbital periods.
}
  \label{Fig03}
\end{figure*}

In Fig.~\ref{Fig03} we show 
other present-day SySt properties 
(eccentricity, 
semi-major axis,
WD and red giant masses,
accretion rate
and
accretion-powered WD luminosity)
and compare the
distributions obtained in
isolated binary evolution
(i.e. without dynamics)
and
inside our GC models.
As some of these properties
might change during
the SySt life-time, we show
the properties at the moment
the accretion-powered 
WD luminosity is maximum.

With respect to the main orbital elements,
we can clearly see that differences with
respect to SySts formed in isolation and
those in GCs. 
In particular, GCs host relatively
more systems with circular orbits and
smaller semi-major axis.
This is another illustration of the
above-mentioned role played by dynamics
in shaping the SySt properties.
Additionally, we can see that the
eccentricity distribution is roughly
bimodal, in which we see the
binaries that managed to circularize 
and the wider binaries that peak 
near 
{$\sim0.4-0.6$}.
Moreover, most SySts have semi-major
axis ranging from a few au up to
$\sim100$~au.

Concerning the component masses,
not surprisingly, all WDs are
carbon-oxygen, as they are formed
similarly to single stars, i.e.
without Roche-lobe overfilling.
Additionally, 
most of them have masses between
$\sim0.55$ and $\sim0.9$~\Msun, 
but a few are more
massive than the Sun.
The red giant masses are mostly concentrated
between $\sim0.7$ and $\sim0.9$~\Msun,
which is directly connected with the
main-sequence turn-off for the 
metallicity and present-day time
assumed here.
However, some  have masses smaller
than $\sim0.6$~\Msun, which is due to 
the strong mass loss through winds
before reaching the SySt phase of 
maximum accretion-powered WD luminosity.
Regarding the evolutionary status of 
the red giant donor, we found that
most ($\gtrsim80$~per~cent) belong to
the first giant branch, while the
remaining are mostly thermally-pulsing
asymptotic giant branch stars.

The accretion rates onto the WD
range from
$\sim10^{-8}$
to
$\sim10^{-7}$~\Msun~yr$^{-1}$,
but most concentrate around
$10^{-7.5}$~\Msun~yr$^{-1}$.
Such rates are high enough so
that thermonuclear burning 
of the accreted material
on the WD surface occurs,
either steadily or unstably.
According to the 
\citet{Nomoto_2007}
criterion,
$\sim25$~per~cent of the
systems reached the phase of
stable hydrogen burning,
while the remaining are likely
symbiotic (recurrent) novae.
Concerning the accretion-powered 
WD luminosity,
the distribution is much broader than the
accretion rate one, but limited to values 
between 
$10$ and $\sim300$~\Lsun.
The lower limit comes from our definition
of SySt and the upper limit is a direct
consequence of the accretion rates
coupled with the WD properties.
We would like to stress that such 
WD luminosities
are basically lower limits, 
as we do not include
in our modelling computations of 
nuclear-powered luminosities, i.e. 
luminosities powered by 
thermonuclear hydrogen burning, 
which provides WD luminosities 
greater than 
$\sim10^3$~\Lsun
\citep[e.g.][]{Nomoto_2007}.
Indeed, most known SySts have WD 
luminosities
$\gtrsim10^3$~\Lsun,
which cannot be explained solely
by accretion
\citep{Mikolajewska_2010}.

The last property we discuss is the
SySt life-times. 
Most systems spend $\sim0.5-2$~Myr 
in the SySt phase. 
These SySt life-times
are likely due to their age
(they are ${\sim11-12}$~Gyr old),
coupled with their red giant masses
(they are close to the turn-off
mass, which is ${\sim0.8-0.9}$~\Msun) 
and the low mass loss rate 
(due to the low metallicity).
The red giant phase in these systems
is much longer, but only in a fraction 
of the red giant life the 
accretion-powered WD luminosity 
is ${\gtrsim10}$~\Lsun, 
which is our condition for the 
occurrence of the symbiotic phenomenon.

While taking into account all
properties together, we can see that
properties of GC SySts and of those formed
in isolation are rather similar, with the
exception of the
orbital period,
semi-major axis,
and
eccentricity.
In the parameter space comprised by 
these properties, the region from which
GC SySts come is considerably smaller
than that from which isolated SySts
come.
This is due to role played by dynamics
in destroying SySt progenitors and
reducing in turn the region in the
parameter space.
Interestingly, this is quite the opposite
of what happens with 
cataclysmic variables in GCs, in 
which dynamics extend the region in the
initial parameter space from where 
they come
\citep[e.g.][]
{Belloni_2016a,
Belloni_2017a,
Belloni_2017b,
Belloni_2019}.

\section{Why not a single 
symbiotic star has ever 
been confirmed?}
\label{resultsGC}

We have just seen that SySt
formed through the 
wind-accretion channel
have very long orbital 
periods ($P\gtrsim10^3$~d)
and most have initial orbital 
periods longer than those 
defining the initial hard-soft
boundaries in our models
(see Fig.~\ref{Fig01}).
Therefore, we do expect that
most SySt would be destroyed 
during the GC evolution.
However, binaries with such long
orbital periods could in principle
still survive in less dense GCs, 
especially if they are beyond the 
half-mass radius, residing
in the GC outskirts.
In this way, we shall investigate the
physical reasons for the observational
lack of SySts in real GCs.

\subsection{Dynamical Destruction}

\begin{figure}
\begin{center}
\includegraphics[width=0.975\linewidth]
{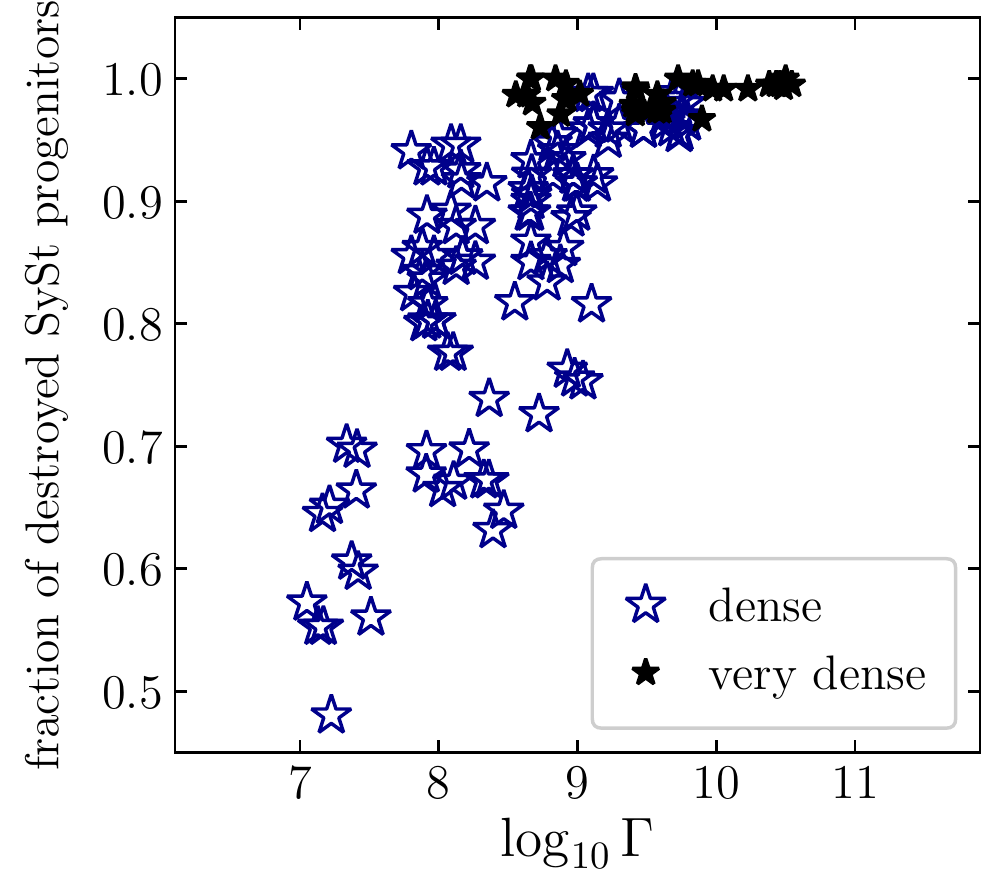} 
\end{center}
\caption{Fraction of destroyed
SySt progenitors as a
function of the initial GC
stellar encounter rate ($\Gamma$).
Filled stars correspond to more
realistic models, according to
the radius--mass relation found by
\citet{Marks_2012}, while
open stars to the remaining models.
Notice the clear correlation between
those two quantities and
the extremely high fractions found
amongst our models, especially for those
very dense.
}
  \label{Fig04}
\end{figure}

In Fig.~\ref{Fig04}, we show the fraction
of destroyed SySt progenitors 
as a function of the 
\textit{initial} 
GC stellar encounter rate, 
given by 
{${\Gamma=\rho_0^2 r_c^3 \sigma_0^{-1}}$}
\citep{Pooley_2006},
where 
$\rho_0$, 
$r_c$,
and 
$\sigma_0$ are 
the central density,
the core radius,
and 
the mass-weighted 
central velocity dispersion,
respectively.
We note that $\Gamma$ is a somewhat
better indicator of the strength of 
dynamics one would expect during the 
GC evolution than individual quantities, 
e.g. the initial
central density, initial concentration,
etc.
In the figure, we separate the clusters
according to their concentration.
\textit{Very dense} models roughly
follow the \citet{Marks_2012}
radius--mass relation, i.e. 
models with initial half-mass radii 
of $\approx1.2$~pc, which are likely
more realistic models.
This is because this relation is in good
agreement with the observed density 
of 
molecular cloud clumps, 
star-forming regions 
and 
globular clusters,
and provides
dynamical evolutionary time-scales for
embedded clusters consistent 
with the life-time of ultra-compact 
H\,{\sc ii} regions and the time-scale 
needed for gas expulsion to be active 
in observed very young clusters, as 
based on their dynamical modelling
\citep[e.g.][and references therein]{Belloni_2018a}.
\textit{Dense} models comprise the remaining
clusters, which are still dense, but
somewhat less dense.

\begin{figure*}
\begin{center}
\includegraphics[width=0.975\linewidth]
{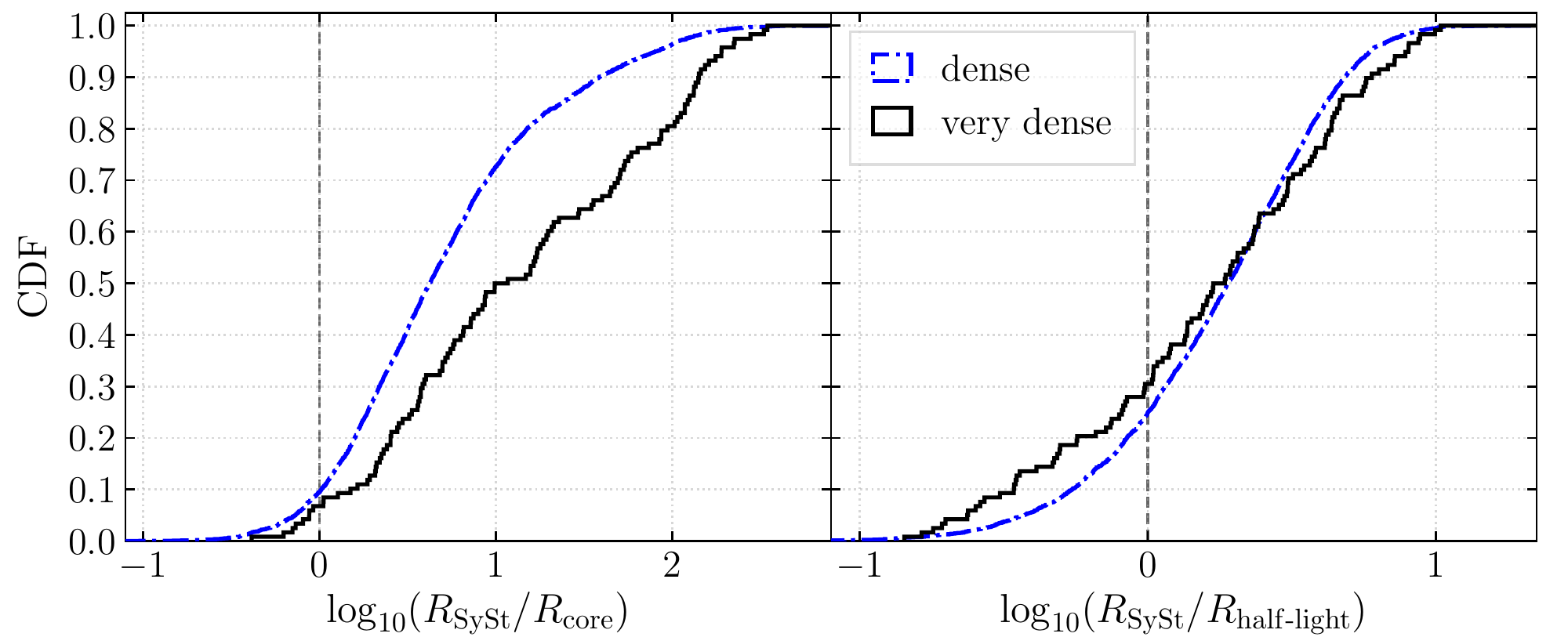} 
\end{center}
\caption{Cumulative radial distribution 
function (CDF) for present-day SySts
in our models, with respect to the
cluster 
core radii (left-hand panel) 
and
half-light radii (right-hand panel),
in which models are separated 
according to the initial concentration,
very dense models being those closer
to the
\citet{Marks_2012}
radius--mass relation.
Notice that nearly all SySts
($\gtrsim90$~per~cent) 
are outside the core radius.
Additionally, most 
($\gtrsim70$~per~cent) 
SySts are beyond the half-light radii, 
which implies that, to have more chances
to identify them, observations
should cover more or less the whole GC, 
or, at least, regions far beyond the 
half-light radii.
This also illustrates
how difficult is 
to detect them, given the extended 
areas of GCs.
}
  \label{Fig05}
\end{figure*}

From Fig.~\ref{Fig04},
we can see that there is a clear 
correlation between $\Gamma$ and
the fraction of destroyed 
SySt progenitors,
which is similar to what has been
found for their relatives 
cataclysmic variables 
\citep{Belloni_2019}.
We carried out 
Spearman's rank correlation tests,
and found a strong correlation 
for the dense models ($r=0.75$)
with more than $99.99$~per~cent confidence.
However, for the very dense models,
the test suggests no correlation,
as these models have naturally 
high $\Gamma$ and very high 
destruction rates.
The fractions of
SySt progenitors that are destroyed
in dense clusters are huge
($\sim84.4\pm12.3$~per~cent, on average), 
which clearly
suggests that the mass density of
SySts are much lower (if not
negligible) in GCs than in 
non-crowded environments.
For those very dense, which
are supposedly more realistic
GC models, the fraction of destroyed
SySts increases to impressive
$\sim98.7\pm1.1$~per~cent.

Therefore, should GCs be born
as dense as proposed by
\citet{Marks_2012},
our results provide 
a natural explanation for the
lack of SySts in GCs,
which is due to dynamics playing
an extremely important role in
destroying their progenitors.
However, if GCs are not
that dense initially, then 
dynamical destruction of SySts
alone would not explain their 
absence in GCs.
If so, we could still expect
non-negligible numbers of
SySts in GCs and there should be
additional reasons for the fact
that not a single one has been 
discovered so far.

\subsection{Spatial Distribution}

In Fig.~\ref{Fig05}, we depict
the cumulative radial distribution 
function for present-day SySts
in all our models, with respect to 
the cluster core radii (left-hand panel) 
and
half-light radii (right-hand panel).
As before, models are separated 
according to the initial concentration,
very dense models being those closer
to the
\citet{Marks_2012}
radius--mass relation.
From the SySt spatial distribution, 
we can clearly see that the
overwhelming majority of systems
are far from the central parts.
Additionally, given their long initial 
orbital periods, it is not surprising
that they only managed to survive in
(very) dense GC environments because 
of that.

Considering all models, we found that
most ($\gtrsim70$~per~cent)
are beyond the half-light radii 
and
nearly all ($\gtrsim90$~per~cent)
are beyond the core radii.
Those models in which SySts survive inside
(or nearly) the GC cores
are characterized by large cores
($\gtrsim3$~pc),
which provides that, albeit rare,
the core relaxation time might
be still sufficiently long, preventing
in turn very frequent and strong
dynamical disruptive interactions.
Indeed, long relaxation time means 
small density and, then, low 
number of dynamical interactions.

In the left-hand panel of Fig.~\ref{Fig05}, 
we can see that SySts are more 
centrally concentrated in the dense 
clusters than in the very dense ones.
This is because SySts and their progenitors 
are more massive than the average stars 
inside the GCs. Thus, due to mass segregation,
they sink, on a time-scale proportional 
(shorter by the ratio between
the average mass and the SySt mass) 
to the half-mass relaxation time,
towards the central parts.
Moreover, in these less dense clusters,
the probability for dynamical interaction, 
and in turn for binary (SySt progenitors) 
dynamical disruption, is smaller than in 
the very dense models.
This provides better chances for the 
SySts to survive the mass segregation 
process and disruptive dynamical 
interactions.
Indeed,
for the very dense models, 
the mass segregation time-scale is
shorter than for the dense models.
So,
SySt progenitors sink faster and 
are quicker destroyed in the very 
dense models.
In the right-hand panel of Fig.~\ref{Fig05},
the relation between the distributions 
of dense and very dense models seem to be 
different than that in the left-hand panel.
This apparent difference is caused
by the non-linear interplay 
between core and half-mass radii 
amongst different cluster models.

Most importantly regarding observations, 
since most systems are
found beyond the half-light radii,
the spatial distribution of SySts in
GCs implies that one would need at least 
extended observations so that the 
SySt population could be recovered.
In particular, a good coverage of the
GC outskirts seems to be crucial to have
any real chances to identify them.

\subsection{Expected Number}

We have shown previously that dynamics
play a significantly important role in
destroying SySt progenitors, especially
in very dense clusters.
Despite that, we also showed that some 
SySts are still expected to exist
in GCs at the present day,
which are not destroyed because the less
crowded region to which they belong.
This is intrinsically connected with the
half-mass relaxation time, which should 
be long enough so that SySts from the 
cluster halo will not have time to mass 
segregate and be destroyed in dynamical 
interactions.
Provided these two facts, one might still
wonder why we fail to observationally detect
these systems.
We provide in what follows additional
arguments for that, which are based on
the expected number of SySts in GCs.

We have considered that the present day is
somewhere between $11$ and $12$~Gyr, 
which corresponds to the expected cluster
ages in the Galactic GC population.
Within these time interval, we found that
the SySt formation rate is roughly uniform
when taking into account all our 144 GC models.
This provides, on average, 
a SySt formation rate given by
$\sim4.22\pm0.63$ SySt per Myr 
in the whole sample of GC models.
Given such a uniform formation rate 
and the total number of GC models, 
we have, on average,
a birth rate of 
$\sim0.0293\pm0.0044$ SySt per Myr per GC.
If we optimistically assume that the SySt 
life-time is one Myr, then we would expect 
to be able to observe this amount of SySt 
in a GC within an Myr.
This then provides that we would expect 
a probability of detection, in observations 
taken in the past $\sim100$~yr, to be 
$10^{-4}$ times the formation rate in 
Myr$^{-1}$, which gives
$\sim2.9\times10^{-6}$.

We would like to stress that this estimate 
is based on ideal situations in which 
we would be able to detect the SySt 
with $100$~per~cent confidence, 
which in reality is not the case.
In addition, the SySt life-time could
be shorter than an Myr in a significant
fraction (or even most systems) 
of the population.
Thus, this probability of
$\sim2.9\times10^{-6}$
should be interpreted 
as an upper limit, and a more realistic 
detection probability would be smaller 
than this.
Therefore, albeit not impossible,
it is rather unlikely that we 
would be able to detect any SySt in the 
Galactic GC population, provided the low 
occurrence of SySts in these stellar systems.

\section{Best cluster targets 
for future observations}

\begin{figure*}
\begin{center}
\includegraphics[width=0.99\linewidth]
{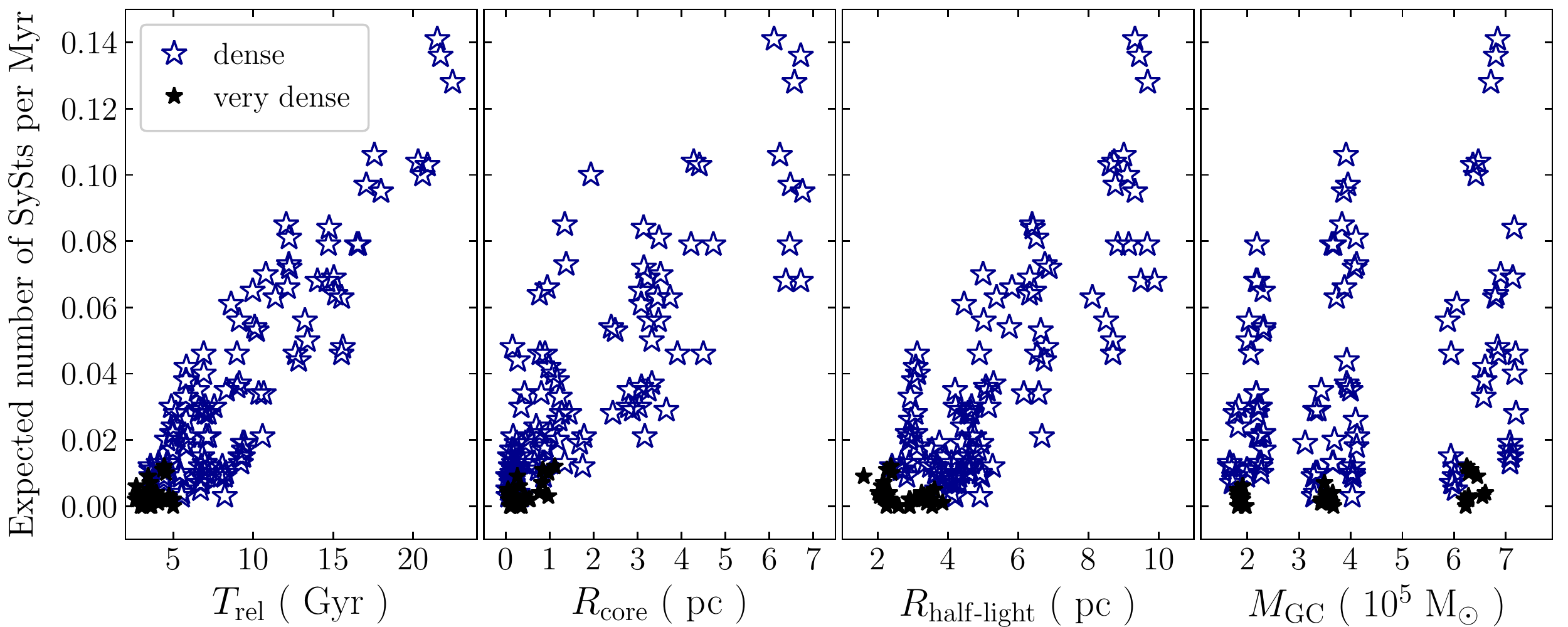} 
\end{center}
\caption{Average present-day SySt abundance within an Myr 
as a function of some GC properties, namely
half-mass relaxation time
($T_{\rm rel}$, first panel),
core radius
($R_{\rm core}$, second panel),
half-light radius
($R_{\rm half{\text -}light}$, second panel),
and
total mass
($M_{\rm GC}$, fourth panel).
Very dense models are the ones closer to the
\citet{Marks_2012} radius--mass 
relation, and dense the remaining
models.
There are clear correlations in the
first three panels, while no 
(or very weak, if at all) in the
last panel.
This suggest that the ideal targets to
search for SySts are 
relatively sparse clusters
with sufficiently long 
half-mass relaxation times.
}
  \label{Fig07}
\end{figure*}

In Fig.~\ref{Fig07}, we present the
expected number of SySts within an
Myr against a few 
present-day GC properties,
namely 
half-mass relaxation time
(first panel),
core radius
(second panel),
half-light radius
(third panel),
and
total mass
(fourth panel).
As before, we separate the models
according to its initial concentration,
very dense being those closer to the
\citet{Marks_2012} radius--mass 
relation.
We can see a clear
correlation between the expected
number of SySts and the 
half-mass relaxation time,
core and half-light radii.
Indeed, we carried out 
Spearman's rank correlation tests
and found a strong correlation 
with more than $99.99$~per~cent confidence, 
in all cases, being the rank values
given by 
$\approx0.77$,
$\approx0.82$,
and
$\approx0.71$,
respectively.
On the other hand, there is
apparently no 
(or very weak, if at all) 
correlation between the expected
number of SySts and the total GC mass, 
which is confirmed by 
the correlation test that
provides a rank value of
$\approx0.26$ with at least
$99.29$~per~cent confidence.
This suggests that the best GC targets
are those relatively extended clusters
with relatively long half-mass
relaxation times.

Regarding correlations amongst GC properties, 
there is a clear correlation
between 
their Galactocentric distances 
and 
their half-light radii
\citep{vandenBergh_1991,Baumgardt_2018},
which is likely due to the strong
tidal fields in the inner parts of the 
Milky Way.
Moreover, there is a clear 
observational correlation
between the half-mass
relaxation times and the half-mass radius
\citep{Baumgardt_2018},
which is not surprising 
since the
half-mass relaxation time is 
proportional to
$R_{\rm half{\text -}mass}^{1.5}$.
These correlations indicate that the 
best GC targets should also be relatively 
far from the Milky Way centre and corroborate
our finds discussed previously.

At this point,
we are able to answer the
question regarding the properties best
GCs should have to be considered ideal
targets to search for SySts.
{\it
One should search for
SySts in the outskirts
of
nearby 
low-density clusters 
(given their 
large 
radii,
angular size,
and
brightness) 
whose half-mass relaxation times
are considerably long and their locations
are not so close to the Galactic Centre.
}

Within the catalogue by
\citet{Baumgardt_2018},
nearby low-density clusters with 
relatively long
half-mass relaxation times and 
relatively large
Galactocentric distances are,
e.g.:
NGC~288,
NGC~4372,
NGC~4590,
NGC~4833,
NGC~5897,
NGC~6362,
NGC~6809,
and
Pal~11.
All these clusters have 
half-mass relaxation times
$\gtrsim3$~Gyr,
central densities
$\lesssim400$~\Msun~pc$^{-3}$,
distances
$\lesssim12$~kpc,
projected half-light radii
$\gtrsim5$~pc,
and
Galactocentric distances
$\gtrsim5$~kpc.

The clusters investigated in the
\muse~survey are in general relatively
dense, many being core-collapsed, and
only the central parts have been covered 
(i.e. up to the half-light radii),
which are usually preferred 
because there is less confusion regarding
the GC membership.
Only one cluster similar to those listed
above was investigated with \muse,
namely NGC~3201
\citep{
Kamann_2018,
Giesers_2018,
Giesers_2019,
Gottgens_2019b}.
However, the pointings for this cluster
covered basically the central parts, 
well inside the half-light radius.
Despite these authors investigated
binaries in detail, by providing the
orbital period and eccentricity
distributions (mainly for main-sequence 
binaries) for the first time in a
GC, they could not find any cataclysmic
variable nor SySt.
Perhaps, if there were pointings
in regions farther from the central parts,
some interesting accreting WD binaries
could be recovered, including the 
long-period ones, such as SySts.

Another interesting cluster investigated
with \muse~is NGC~6656, which possibly
harbours a nova remnant that could have 
originated in a symbiotic nova, instead 
of a classical nova
\citep{Gottgens_2019a}.
This nova remnant lies within the core radius, 
which provides a rather low probability 
for long-period systems such as SySts to 
survive.
From our results, $\sim10$~per~cent of the
predicted SySts are inside the core radii
of our models.
So, albeit unlikely to find them there,
it is not impossible.
Additionally, NGC~6656 has one of the
largest cores in the Galactic GC
population \citep[][2010 edition]{Harris_1996},
so it is rather consistent with our results that 
such a type of GC 
might harbour an SySt within its core.
Therefore, 
it is definitely worthwhile to put 
more observational efforts on this source 
to disentangle the possibility that it 
could be a symbiotic nova remnant.

Finally, we mention $\omega$~Cen and 47~Tuc as promising clusters to harbour SySts.
$\omega$~Cen is an extended low-density cluster,
characterized by a long half-mass relaxation time and large half-light and core radii.
47~Tuc, on the other hand, is one of the Galactic GCs with the largest stellar interaction rates \citep[][]{Bahramian_2013,Cheng_2018}, due to its small and very dynamically active core.
This likely explains the lack of SySt candidate detections in the central parts of this cluster, where most of the searches for interacting binaries have been performed so far
\citep[e.g.][]{Edmonds_2003a,Edmonds_2003b,Knigge_2008,Rivera_2018,Campos_2018}.
However, 47~Tuc is a non-core-collapsed cluster and, given its relatively long half-mass relaxation time and size, this cluster as a whole cannot be considered dynamically old.
For this reason, it is a good candidate to look for SySts in its outer parts.

\section{Summary and Conclusions}

We refined here the nature of SOPS~IV~e-94,
the promising SySt candidate in 
$\omega$~Cen,
by obtaining a deep SALT optical spectrum
and concluded that this object cannot 
be classified as an SySt.
This is because 
none of the features typical for 
an SySt are present in the
spectrum of SOPS~IV~e-94,
e.g. 
there is no trace of any 
emission lines, including the 
strongest H$\alpha$ emission 
that is always visible in SySts.

We investigated SySts
formed through the wind-accretion channel
in 144 globular cluster models evolved
with the \mocca~code with the aim of 
explaining why not a single one has
ever been identified in a Galactic
globular cluster.

We found that most progenitors
of these systems are destroyed
in dense globular clusters before 
effectively becoming SySts
at the present day.
This happens because the progenitors
of these systems have initially orbital
periods ($\gtrsim10^3$\,d)
that are comparable to 
(or even much longer than)  
the orbital period separating soft 
from hard binaries in the clusters.
This puts them into the group of
soft binaries and makes their 
destruction through dynamical 
interactions sufficiently
easy over the cluster evolution
time-scale
($\sim11-12$\,Gyr).

However, in less dense clusters,
SySts should still be present.
Most of these SySts
($\gtrsim70$~per~cent)
are located 
far from the 
cluster central parts 
and 
beyond the 
cluster half-light radii
(i.e. less dense regions),
which is the main reason 
why they managed to survive
in the clusters.
This also makes their detection
difficult, provided the large areas 
of the globular cluster outskirts.
Additionally,
given the typical life-times
of SySts
($\sim1$~Myr),
their expected numbers are
extremely low 
($\lesssim1$~per~globular~cluster~per~Myr).

Our results provide therefore an
explanation for the observed absence 
of SySts in 
Galactic globular clusters, which
occurs due to
a combination of 
three important effects: 
(i) most are destroyed in dynamical interactions, and most that survived (ii) are far from the central parts and (iii) are sufficiently rare, which makes their discovery in current dedicated observational surveys rather difficult.

Coupling the 
properties of 
SySts
and 
globular clusters, 
we found that
the best chances to identify them 
are in the outskirts of
nearby low-density clusters 
with relatively long 
half-mass relaxation times
and relatively large 
Galactocentric distances,
by means of
either 
high-quality spectroscopy 
or 
photometry using narrow-band 
filters centred on 
either
the \mbox{He\,{\sc ii}} and H$\alpha$ 
or 
the Raman-scattered 
\mbox{O\,\sc{ vi}}
emission lines.

Since the majority of known
SySts are formed through
the Roche-lobe overflow channel
(not addressed here), 
it remains to be shown that
the absence of whole population
of SySts in globular clusters might 
be explained in a similar fashion
to what we presented here. 
In follow-up works, we intend to investigate
these systems not only in 
Galactic globular clusters but also in
non-crowded fields of our Galaxy
and of other galaxies.

\section*{Acknowledgements}

We would like to thank Michael M. Shara for his feedback on this paper.
We also thank an anonymous referee for the comments and suggestions that helped to improve this manuscript.
We thank MCTIC/FINEP (CT-INFRA grant 0112052700) and the Embrace Space Weather Program for the computing facilities at the National Institute for Space Research, Brazil.
This paper is based on spectroscopic observations made with the Southern African Large Telescope (SALT) under programme 2019-2-SCI-021 (PI: K. I{{\l}lkiewicz). Polish participation in SALT is funded by grant No. MNiSW DIR/WK/2016/07.
DB was supported by the grants {\#2017/14289-3} and {\#2018/23562-8}, S\~ao Paulo Research Foundation (FAPESP).
This research has been partly funded by the National Science Centre, Poland, through grant OPUS 2017/27/B/ST9/01940 to JM.
%
%
MRS acknowledges financial support from FONDECYT grant number 1181404. 
MG and DB was partially supported by the Polish National Science Center (NCN) through the grant UMO-2016/23/B/ST9/02732. 
LERS thanks NASA for support under grant 80NSSC17K0334.
CVR would like to thank funding support from Funda\c c\~ao de Amparo \`a Pesquisa do Estado de S\~ao Paulo (FAPESP, Proc. 2013/26258-4) and CNPq (Proc. 303444/2018-5).

\section*{Data availability}

The data underlying this article can be obtained upon request to Mirek Giersz (mig@camk.edu.pl) and after agreeing to the terms of the \mocca~License. The license can be found in \href{https://moccacode.net/license/}{\texttt{https://moccacode.net/license/}}.


\bibliographystyle{mnras}
\bibliography{references} 



\bsp	
\label{lastpage}

\end{document}